\documentclass[review,3p,authoryear]{elsarticle}

\usepackage{amsmath,amssymb,amsthm,mathtools}
\usepackage{setspace} 
\usepackage{booktabs,graphicx,array,longtable,threeparttable}
\usepackage{xcolor,float,microtype,needspace}
\usepackage{caption,ragged2e}
\usepackage{hyperref}

\hypersetup{
    colorlinks = true,
    linkcolor  = blue,
    citecolor  = blue,
    urlcolor   = blue,
    filecolor  = black,
    pdfborder  = {0 0 0}
}

\definecolor{Navy}{HTML}{294B68}
\definecolor{Teal}{HTML}{087F8C}

\newtheorem{theorem}{Theorem}[section]
\newtheorem{lemma}[theorem]{Lemma}

\theoremstyle{definition}
\newtheorem{definition}[theorem]{Definition}

\newcommand{\F}{\mathcal F}
\newcommand{\A}{\mathcal A}
\newcommand{\C}{\mathcal C}

\newcommand{\ket}[1]{\lvert #1\rangle}
\newcommand{\bra}[1]{\langle #1\rvert}

\DeclareMathOperator{\CX}{CX}
\DeclareMathOperator{\NOR}{NOR}
\DeclareMathOperator{\supp}{supp}

\newcommand{\GF}{\mathbb F_2}

\newcommand{\ii}{\mathrm i}

\makeatletter
\def\tnotemark[#1]{%
  \@for\mytmark:=#1\do{%
    \expandafter\ifcase\elsRef{\mytmark}\or
      $^{\dagger}$\or $^{,\dagger\dagger}$\fi
  }%
}
\def\tnotetext[#1]#2{%
  \g@addto@macro\@tnotes{%
    \refstepcounter{tnote}%
    \immediate\write\@auxout{%
      \string\Newlabel{#1}{\thetnote}}%
    \def\thefootnote{%
      \ifcase\c@tnote\or$\dagger$\or
      $\dagger\dagger$\fi}%
    \footnotetext{#2}%
  }%
}
\makeatother

\graphicspath{{figures/}{./}}

\newcommand{\Figure}[2][0.95\textwidth]{%
    \IfFileExists{figures/#2.pdf}{%
        \includegraphics[width=#1]{figures/#2.pdf}%
    }{%
        \IfFileExists{#2.pdf}{%
            \includegraphics[width=#1]{#2.pdf}%
        }{%
            \fbox{%
                \parbox[c][28mm][c]{.86\linewidth}{%
                    \centering\small
                    Missing source figure: \texttt{\detokenize{#2.pdf}}
                }%
            }%
        }%
    }%
}

\newcommand{\FigureNote}[1]{%
    \par\vspace{3pt}\noindent
    \begin{minipage}{0.95\textwidth}
        \scriptsize
        \justifying
        \setlength{\parindent}{0pt}%
        #1\par
    \end{minipage}%
}

\newcommand{\LongTableNote}[1]{%
    \par\vspace{3pt}\noindent
    \begin{minipage}{\textwidth}
        \scriptsize
        \justifying
        \setlength{\parindent}{0pt}%
        #1\par
    \end{minipage}%
}

\begin{document}
\doublespacing
\begin{frontmatter}

\title{Constraint-Preserving Quantum Circuit Compilation for Graph Independent Sets via Persistent Occupancy Encoding\tnoteref{code1}}

\tnotetext[code1]{
Code and experimental data are publicly available at
\url{https://github.com/Moonooo24/persistent-occupancy-compiler}.
Xiaoxing Liu acknowledges financial support from the Major Program
of the National Social Science Fund of China (Grant No. 24\&ZD117).
We thank the CMU COLOR maintainers and contributors for providing
benchmark graphs, the Qiskit developers for circuit synthesis
and routing tools, the QuCLEAR authors for their baseline
implementation, and the Cirq developers for tools supporting
global-phase verification.
}

\author[1,3]{Zhiyi Wang\corref{cor1}}

\author[1,3]{Xiaorui Xue}

\author[1,2,3]{Xiaoxing Liu}

\cortext[cor1]{Corresponding author. Email address: wangzhiyi924@gmail.com}

\affiliation[1]{
    organization={School of Cyber Science and Engineering, Southeast University},
    addressline={2 Southeast University Road, Jiangning District},
    city={Nanjing},
    postcode={211189},
    country={China}
}

\affiliation[2]{
    organization={School of Economics and Management, Southeast University},
    addressline={2 Southeast University Road, Jiangning District},
    city={Nanjing},
    postcode={211189},
    country={China}
}

\affiliation[3]{
    organization={Laboratory of Philosophy and Social Sciences at Universities in Jiangsu Province-Fintech and Big Data Laboratory of Southeast University},
    addressline={2 Southeast University Road, Jiangning District},
    city={Nanjing},
    postcode={211189},
    country={China}
}
\begin{abstract}
Constraint-preserving mixers for graph independent sets repeatedly evaluate overlapping neighborhood conditions, increasing CX costs across ordered quantum optimization circuits. We propose a structure-aware compiler that reuses these conditions through persistent occupancy encoding. Exploiting mutual exclusion within true-twin classes, our reversible encoding stores class occupancy on the original data register across mixer operations and cost layers. Local mixers are synthesized using target compensation, blockwise parity queries, and coherent restoration; nonuniform cost phases are preserved by an exact coefficient transformation. We prove that the compiled circuit reproduces the original program's unitary action on the feasible subspace for any prescribed operation order and independent layer parameters, and establish exact separability of source-level caching costs under a fixed synthesis recipe. Across 134 evaluation programs and 72 independent confirmation programs, total CX counts fall by 22.61\% and 19.87\%, respectively, relative to the best of six reference implementations selected per program. Against a matched no-cache implementation, persistent encoding yields additional reductions of 4.10\% and 9.13\%.
\end{abstract}
\end{frontmatter}
\doublespacing 

\section{Introduction}
For quantum combinatorial optimization problems with hard constraints, the mixer must preserve the feasible subspace so that the evolution remains within it. The quantum alternating operator framework uses constraint-preserving mixers to implement transitions within the feasible subspace~\citep{hadfield}. In the graph independent set problem, for example, a local mixer operation at a vertex is controlled by the condition that all its neighbors are unoccupied, and the control conditions for different target vertices may involve overlapping sets of neighbors.
Implementing local mixing directly by computing control predicates, applying controlled rotations, and uncomputing those predicates repeatedly evaluates overlapping neighborhood conditions, causing the two-qubit gate cost to accumulate over successive mixer sweeps. Because local mixer operations generally do not commute, their execution order determines the overall unitary evolution. We study how to turn overlapping neighborhood information into a circuit representation that can be reused across operations to reduce the CX cost of the complete circuit while preserving the specified operation order and the semantics of the evolution.

Prior work has developed constraint-preserving quantum optimization methods through mixer design, feasible-state preparation, and evolution in restricted subspaces~\citep{farhi14,bartschi20,fuchs22,fuchs24,compact25}. Encodings for discrete optimization and structured synthesis of conditional rotations also provide ways to reduce circuit costs~\citep{dqir,bergholm05,vale}.
In an ordered program comprising multiple local mixer operations and intervening cost layers, overlapping neighborhood conditions create opportunities for reuse across operations. The central task is to construct an occupancy representation that is updated in step with legal local transitions, so that shared conditions remain available throughout the ordered evolution while preserving access to the original variables, the semantics of their legal flips, the cost phases, and coherent correctness.

\begin{figure}[H]
\centering
\Figure{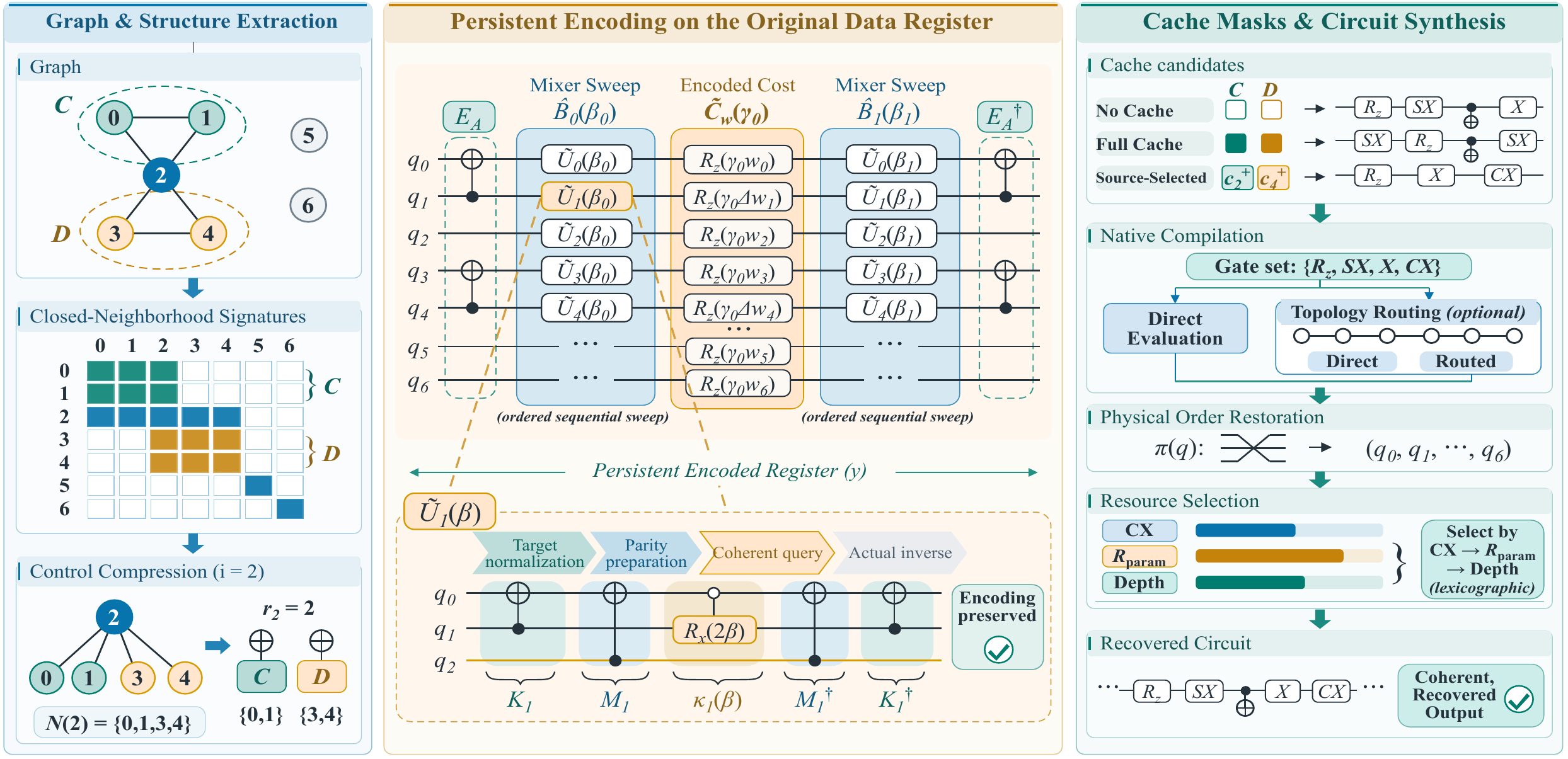}
\caption{Structure-aware compilation pipeline based on persistent occupancy encoding}
\label{fig:overview}

\FigureNote{Closed-neighborhood signatures determine the true-twin equivalence classes and cache candidates. Reversible occupancy encoding on the original data register is maintained across ordered mixer operations and cost layers and decoded at the end of the program. A local mixer macro comprises target normalization, a parity-conditioned query, a rotation kernel, and actual uncomputation. The compiler selects the final circuit from the empty-cache, full-cache, and source-level optimal cache schemes using the resources of the complete native circuit, with optional physical routing.}
\end{figure}

In an undirected graph, vertices with identical closed neighborhoods belong to the same true-twin equivalence class. Its members are pairwise adjacent and have identical external adjacencies~\citep{habibpaul10}. On the independent-set feasible domain, at most one vertex in a class can be selected, so the parity of the binary variables within the class exactly represents its occupancy. We exploit this structure to encode class occupancy information in data coordinates that are continually updated by legal local transitions, allowing overlapping neighborhood conditions to be reused across ordered mixer operations and intervening cost layers. We then construct compiled circuits that preserve unitary action on the feasible subspace through blockwise parity queries, coherent restoration, and a transformation of the cost coefficients. Figure~\ref{fig:overview} shows the compilation pipeline from true-twin class identification to complete circuit selection.

Our experiments use six fixed reference implementations to define a per-program CX benchmark and a matched no-cache control using the same query synthesis mechanism to quantify the independent contribution of persistently maintained occupancy information. The reference implementations cover structured controlled rotations, open and closed CPG--Walsh synthesis, and matched full-space exact kernel implementations~\citep{compact25,vale,bergholm05,mottonen,welch}.
Relative to the lowest reference CX count per program, the total CX count decreases by 22.61\% in the main evaluation set of 62 distinct labeled graphs and 134 ordered programs. In the independent confirmation set of 36 nonisomorphic graphs and 72 programs, it decreases by 19.87\%. Every program in both sets achieves a CX reduction. Matched ablations also show that persistent occupancy encoding further reduces the total CX counts by 4.10\% and 9.13\%, respectively, under the same query synthesis mechanism. After physical routing using a SABRE configuration~\citep{li19sabre} and restoration of the original wire order, the total CX counts decrease by 9.14\% on the linear topology and by 12.50\% on the two-dimensional grid.

The main contributions of this paper are as follows.
\begin{enumerate}
\item \textbf{Persistent occupancy encoding for the graph independent-set feasible domain.} We turn mutual exclusivity of occupancy within true-twin classes into a shared data representation that can be continually updated. We construct a reversible encoding on the original data register and use target compensation to preserve the semantics of accessing and legally flipping the original variables. The encoded representation persists across ordered mixer operations and intervening cost layers, and an exact coefficient transformation implements the phase action of costs with nonuniform weights.
\item \textbf{Coherently composable local queries and rotation circuits.} We exploit true-twin structure to organize blockwise parity queries, construct open Gray--Walsh circuits and conditional phase query kernels, and establish a unified coherent restoration contract. We prove that the local mixer macros and their ordered compositions realize the unitary action of the original program on the feasible subspace.
\item \textbf{Source-level cost separability and end-to-end resource selection.} We prove that, under a fixed circuit recipe, source-level caching cost is exactly separable with respect to cache class selection. We derive an amortization criterion for multiple mixer sweeps and reduce caching decisions to per-class marginal cost evaluation. We quantify the CX resource savings through complete native synthesis, matched ablations, independent confirmation, and physical routing.
\end{enumerate}

The rest of this paper is organized as follows. Section~\ref{sec:related} reviews related work, Section~\ref{sec:spec} specifies the compilation task and establishes coherent correctness conditions, Section~\ref{sec:compiler} presents the compilation method, Section~\ref{sec:evaluation} describes the experimental setup and performance evaluation, and Section~\ref{sec:con} concludes.

\section{Related Work}
\label{sec:related}

\subsection{Constraint-Preserving Mixers and Restricted Evolution}

For combinatorial optimization problems with hard constraints, the QAOA mixer layer must connect feasible solutions while preventing amplitude leakage into infeasible states~\citep{farhi14}. Within the quantum alternating operator framework, \citet{hadfield} constructed vertex rotations for maximum independent set controlled by the condition that all neighboring bits are zero.
Local rotations at adjacent vertices generally do not commute, so their execution order is part of the definition of the mixer-layer evolution. \citet{bartschi20} use a Grover-type mixer, shifting part of the implementation complexity to the preparation of a uniform superposition of feasible states. These two approaches implement constraint-preserving mixing through local feasible transitions and global feasible-state preparation, respectively.

The circuit implementation of a mixer also directly affects the resource cost of restricted evolution. \citet{fuchs22} studied the construction of constraint-preserving mixers and their decomposition into basis gates. \citet{fuchs24} used the stabilizer formalism to define logical $X$ mixers on feasible subspaces and analyzed their CX requirements. Subsequent work constructed compact circuits for evolution restricted by sparse projectors using basis transformations and projector structure~\citep{compact25}.
These studies provide methods for reducing the implementation cost of restricted evolution. For sequentially executed local mixers, reusing overlapping neighborhood conditions across operations offers further opportunities for resource optimization.

\subsection{Graph-Structured Encoding and Conditional Rotation Synthesis}

The choice of encoding determines the correspondence among optimization variables, constraints, and quantum circuits. \citet{dqir} systematically compared the representational properties and implementation costs of discrete-variable encodings. For graph independent sets, the uniform external adjacency captured by modular decomposition can be used to organize circuits that query shared neighborhood conditions~\citep{habibpaul10}. True-twin vertices with identical closed neighborhoods form a clique and have identical adjacencies to vertices outside the class.
On the independent-set feasible domain, each true-twin class contains at most one selected vertex, so the parity of its bits equals its occupancy. Because distinct classes are either fully adjacent or nonadjacent, the all-zero neighborhood condition required for a local rotation can be expressed in terms of block occupancy information and the target bit, organizing vertex-wise condition queries into blockwise queries with shared structure.

Whether this structure yields circuit savings also depends on how the conditional rotations are synthesized. General unitary decomposition and isometry synthesis provide general-purpose constructions~\citep{shende06,iten16}. Uniformly controlled rotations, Gray ordering, and Walsh expansions provide specialized decompositions for particular control structures~\citep{bergholm05,mottonen,welch}.
The decomposition and resource trade-offs of multi-controlled gates have been studied separately~\citep{vale,tomesh}. Relative-phase Toffoli gates and conditionally clean workspace bits provide further ways to reduce control overhead~\citep{maslov,khattar}. These studies provide several synthesis methods for reducing the circuit cost of conditional rotations, while the joint optimization of local queries and coherent restoration under a persistent encoding remains open to further investigation.

\subsection{Semantics-Driven Quantum Compilation and End-to-End Resource Evaluation}

Circuit optimization across operations can exploit state relationships and gate-sequence structure in quantum programs. \citet{haener} guide circuit optimization using state properties expressed by program assertions, while \citet{hietala21} formally verify the correctness of circuit transformations in VOQC.
\citet{relational} use linear and nonlinear relationships among variables to optimize phase computations, while the QuCLEAR method of \citet{quclear} optimizes gates across operations by extracting and absorbing Clifford operations. These methods eliminate redundant computation by exploiting program semantics or the algebraic structure of circuits, respectively.

The net resource savings from reuse across operations depend on both the repeated computation eliminated and the cost of maintaining a persistent representation. Persistently maintaining an intermediate representation introduces additional overhead for encoding, state updates, coordinate transformations, and query uncomputation. Backend gate cancellation and resynthesis can also change the resource relationships among source-level implementations. Hardware coupling constraints further increase circuit mapping costs. For example, \citet{li19sabre} proposed SABRE, which uses heuristic search to improve qubit mapping and the selection of swap operations, reducing the implementation costs associated with limited device connectivity.
Source-level gate-count analysis thus characterizes computable resource relationships among local recipes, while complete synthesis and physical mapping determine the actual end-to-end circuit savings.

The above studies provide foundations for constraint-preserving evolution, structured query synthesis, program-semantic optimization, and circuit resource evaluation. We further exploit mutual exclusivity within true-twin classes to represent the occupancy information needed by local neighborhoods in an encoding that is continually updated as legal transitions occur. We support ordered mixer programs through coherent queries and cost-coordinate transformations. Source-level cost separability then provides a computational basis for cache class selection, while complete circuit synthesis and physical routing are used to evaluate the end-to-end resource savings from reuse across operations.

\section{Problem Formulation and Coherent Correctness}
\label{sec:spec}
This section formalizes ordered mixer programs and their compilation correctness conditions, uses an occupancy encoding example to illustrate encoding and restoration for local operations, and establishes a coherent composition theorem and the phase restoration conditions for query kernels.

\subsection{Ordered Mixer Programs and Compilation Objective}
Let $G=(V,E_G)$ be a simple undirected graph with $|V|=n\ge1$. The set of feasible bit strings satisfying the independent-set constraints and the corresponding subspace projector are defined by
\begin{equation}
\F=\{x\in\{0,1\}^n:x_u x_v=0\ \text{for all }uv\in E_G\},\qquad
\Pi_\F=\sum_{x\in\F}\ket{x}\bra{x}.
\end{equation}
Let $N(i)$ denote the open neighborhood and let $h_i(x)=\prod_{j\in N(i)}(1-x_j)$. The quantum occupancy operator is $\widehat x_j=(I-Z_j)/2$, and the associated projector for zero-valued neighborhood controls is defined as $h_i(Z)=\prod_{j\in N(i)}\frac{I+Z_j}{2}$. The canonical mixer operation and cost operator are
\begin{equation}
\label{eq:canonical}
 U_i(\beta)=\exp[-\ii\beta X_i h_i(Z)],\qquad
 D_w=-\sum_{j=0}^{n-1}w_j \widehat x_j,\quad C_w(\gamma)=e^{-\ii\gamma D_w}.
\end{equation}
The compiler parses the decimal representations of the input weights as exact rationals $w_j$, with positive weights corresponding to the standard weighted independent-set objective. On computational-basis branches where all neighborhood bits are zero, $U_i$ applies $R_x(2\beta)$. It acts as the identity on all other branches.

Suppose that the input program contains $p\ge1$ complete mixer sweeps, and let $\sigma_\ell$ be the vertex permutation for sweep $\ell$. A mixer sweep executes the local mixer operations of all vertices in the prescribed order. Its unitary operator is
\begin{equation}
B_\ell=U_{\sigma_\ell(n-1)}(\beta_\ell)\cdots U_{\sigma_\ell(0)}(\beta_\ell).
\end{equation}
The unitary operator for the complete program is
\begin{equation}
\label{eq:whole}
 W=B_{p-1} C_w(\gamma_{p-2})B_{p-2}\cdots C_w(\gamma_0)B_0.
\end{equation}
The layer parameters $\beta_\ell$ and $\gamma_\ell$ are independent. For $p=1$, there is no intervening cost layer. The target compiled circuit $\widehat W$ must satisfy the following condition for all real parameter values.
\begin{equation}
\label{eq:correct}
 \widehat W\Pi_\F=W\Pi_\F.
\end{equation}
This condition applies to any feasible input state and continues to hold when both sides are tensored with the identity on an arbitrary external reference system. Compilation uses the original $n$ data qubits and restores the original physical wire order of the logical data after physical routing.

\subsection{Example of Persistent Occupancy Encoding}
Consider a seven-vertex conflict graph with two true-twin classes $C=\{0,1\}$ and $D=\{3,4\}$. The vertices within each class are adjacent, and every vertex in either class is adjacent to vertex 2. There are no edges between the two classes, and vertices 5 and 6 are isolated. Select class $C$ for persistent occupancy encoding, use vertex 0 as its anchor, and set $E=\CX(1\to0)$.
Because feasible states satisfy $x_0x_1=0$, the encoded coordinates satisfy $y_0=x_0\oplus x_1=x_0+x_1$ and $y_1=x_1$. Here, $y_0$ records the occupancy of class $C$, and $y_1$ retains the original variable value of vertex 1.

For the local mixer operation $U_1(\beta_a)$ at vertex 1, the local mixer macro first applies the target compensation $K_1=\CX(1\to0)$ to recover the original coordinates $x_0,x_1$ in the temporary representation and queries the neighborhood condition with vertex 1 as the rotation target. It then applies the exact inverse of the actual preparation circuit to restore the encoded representation.
With the empty-set basis state $|0\rangle\equiv|0^n\rangle$ as input, the single-vertex occupancy branch $|e_1\rangle$ produced by $U_1(\beta_a)$ corresponds to $E|e_1\rangle=|e_0\oplus e_1\rangle$ in the encoded domain. Its encoded coordinates are $(y_0,y_1)=(1,1)$, corresponding to the original variables $(x_0,x_1)=(0,1)$.
This operation keeps the legal flip at vertex 1 consistent with the update of the occupancy flag of class $C$. Subsequent local operations reuse the updated occupancy information.

Next, consider executing the mixer $U_1(\beta_a)$ at vertex 1, an intervening cost layer $C_w(\gamma)$, and the mixer $U_2(\beta_b)$ at vertex 2 in that order. With $w_0=3,w_1=5$, the encoded cost of the $|e_1\rangle$ branch is $3y_0+(5-3)y_1=5$, giving a phase factor of $e^{5\ii\gamma}$.
Because vertices 1 and 2 are adjacent, the subsequent mixer $U_2(\beta_b)$ leaves this branch unchanged and allows the empty-set branch to rotate toward $|e_2\rangle$. Thus, applying the encoding, the two compiled mixer operations with the intervening cost layer, and then the decoding to the empty-set basis state gives
\begin{equation}
\begin{aligned}
&E^\dagger\widetilde U_2(\beta_b)\widetilde C_w(\gamma)\widetilde U_1(\beta_a)E|0\rangle\\
&\quad=\cos\beta_a\cos\beta_b|0\rangle
-\ii\cos\beta_a\sin\beta_b|e_2\rangle
-\ii e^{5\ii\gamma}\sin\beta_a|e_1\rangle.
\end{aligned}
\end{equation}
This example shows that the occupancy update induced by local mixing, the cost phase, and the subsequent neighborhood query can all be implemented in the same persistent encoded representation, with terminal decoding restoring the original data representation.

For subsequent reference, Table~\ref{tab:notation} defines the principal symbols used for occupancy encoding, local queries, and source-level resource analysis.

\begin{table}[H]
\centering
\caption{Principal notation}
\label{tab:notation}

\begin{threeparttable}
\scriptsize
\setlength{\tabcolsep}{4pt}
\renewcommand{\arraystretch}{1.05}

\begin{tabular*}{\textwidth}{@{\extracolsep{\fill}}lp{0.76\textwidth}@{}}
\toprule
Symbol & Meaning\\
\midrule
$\F,\Pi_\F$ & Independent-set feasible domain and its projector\\
$\C,\A$ & Set of true-twin equivalence classes and set of selected cache classes\\
$E_\A,K_i,T_i$ & Persistent occupancy encoding, target compensation, and $T_i=K_iE_\A$\\
$S_{ib},f_{ib},r_i$ & Original neighborhood blocks, parity query rows in temporary coordinates, and control rank\\
$M_i,\kappa_i$ & Parity preparation and complete local rotation kernel\\
$B(\A),F(\A),J(\A)$ & Source-level CX count for one encoding or decoding, source-level CX count for one complete mixer sweep, and source-level cost objective under a fixed recipe\\
\bottomrule
\end{tabular*}

\begin{tablenotes}[flushleft]
\scriptsize
\item Matrix and row-vector operations are over $\GF$. Weighted sums in the cost are over the rationals or reals.
\end{tablenotes}
\end{threeparttable}
\end{table}

\subsection{Local Restoration Contract and Coherent Composition}
\label{sec:interface}
To formalize the updatable representation, let $E$ be a reversible data encoding and let $E\operatorname{im}\Pi_\F$ be the encoded feasible space. For target $i$, choose a reversible preparation transformation $P_i$ and a complete rotation kernel $\kappa_i(\beta)$, with the preparation transformation allowed to introduce temporary data coordinates. The local mixer macro is defined as
\begin{equation}
\label{eq:macro-general}
 \widetilde U_i(\beta)=P_i^\dagger\kappa_i(\beta)P_i.
\end{equation}
\begin{definition}[Query and Restoration Contract]\label{def:contract}
The preparation transformation $P_iE$ maps the original feasible inputs into the admissible input domain $\mathcal Q_i$ of the kernel. On this domain, $\kappa_i$ implements the prescribed two-level rotation on the target qubit while preserving complex amplitudes and phase relationships between admissible branches. Applying the inverse of the actual preparation circuit then restores the nontarget data and borrowed qubits to their specified states.
If a residual unitary $R_i$ is retained, then $R_i$ acts as the identity on the relevant output subspace. Decoding after an enabled flip corresponds to a legal flip of the original variable $i$.
\end{definition}

\begin{theorem}[Coherent Composition]\label{thm:composition}
If every complete macro satisfies
$\widetilde U_i(\beta)E\Pi_\F=EU_i(\beta)\Pi_\F$, and the encoded cost satisfies
$\widetilde C_w(\gamma)E\Pi_\F=EC_w(\gamma)\Pi_\F$, then composing the macros and cost layers in the order specified in Eq.~\eqref{eq:whole} and applying $E$ and $E^\dagger$ at the beginning and end of the program, respectively, yields a complete compiled circuit satisfying the operator identity in Eq.~\eqref{eq:correct}. This conclusion holds for arbitrary $p$, individually prescribed execution orders for the layers, and independent real parameters.
\end{theorem}
\begin{proof}
Each canonical mixer and diagonal cost operator preserves the feasible subspace, so the output of the corresponding encoded macro lies in $E\operatorname{im}\Pi_\F$. Substituting the local identities in the prescribed operation order makes each adjacent product $E^\dagger E$ of encoding and decoding transformations the identity, and the terminal $E^\dagger$ restores the original data representation. The local identities hold as linear operator identities on the entire feasible subspace and therefore continue to hold after tensoring with an arbitrary reference system.
\end{proof}

\subsection{Residual Phases and Complete Rotation Kernels}
The query and restoration contract requires that the complete mixer macro preserve the specified unitary action on the encoded feasible subspace. For adjacent vertices $u,v$, we have $\mathrm{CZ}_{uv}\Pi_\F=\Pi_\F$. Thus, if the residual diagonal operator produced by synthesis acts as the identity on the relevant output subspace, the complete macro may retain it while preserving its unitary action on the feasible subspace.

For the $r$ compressed zero controls, define the admissible input domain of the complete rotation kernel as
\begin{equation}
\label{eq:care}
\mathcal Q_i=\operatorname{span}\{\ket{0}_i\ket{z}:z\in\{0,1\}^r\}
\,+\operatorname{span}\{\ket{1}_i\ket{0^r}\}.
\end{equation}
On the admissible input domain $\mathcal Q_i$, the rotation kernel applies $R_x(2\beta)$ when all controls are zero and acts as the identity on branches with a zero target bit when the control string is nonzero. The specific gate-level construction determines the unitary extension to the other sectors. This action and the residual-phase restoration condition together define correctness for the local query kernel. The surrounding preparation and its actual inverse connect the kernel to the persistent encoded representation.

\section{Structure-Aware Compilation via Persistent Occupancy Encoding}
\label{sec:compiler}
This section constructs true-twin-driven persistent occupancy encodings and neighborhood-condition query circuits, and provides methods for implementing nonuniformly weighted cost layers and for constructing and selecting caching schemes. All corresponding local circuits satisfy the coherent restoration contract of Section~\ref{sec:spec}.

\subsection{Automatic Discovery of Closed-Neighborhood Modules}

Define $u\sim v$ if and only if $N[u]=N[v]$, where $N[u]=N(u)\cup\{u\}$. Represent the closed neighborhood of each vertex as a list (signature) sorted by vertex identifier and group vertices with identical signatures. This yields the true-twin equivalence-class partition $\C$ uniquely determined by closed-neighborhood equivalence. The partition can be constructed by hash grouping the closed-neighborhood signatures. Given signatures already sorted by vertex identifier and hash operations with expected constant time complexity, grouping takes expected time linear in the total length of the signatures.

\begin{lemma}[Structural Properties of Closed-Neighborhood Equivalence Classes]\label{lem:twins}
Each true-twin equivalence class forms a clique, and the edges between any two distinct classes form either a complete bipartite edge set or the empty set.
\end{lemma}
\begin{proof}
Distinct vertices $u,v$ in the same class satisfy $u\in N[v]$ and hence are adjacent. If there is an edge between classes $C,D$, equality of the closed neighborhoods within the two classes implies $uv\in E_G$ for every $u\in C$ and $v\in D$.
\end{proof}

\subsection{Persistent Occupancy Encoding and Target Normalization}
For any selected set of nonsingleton classes $\A\subseteq\C$, take $a_C=\min C$ as the anchor and define
\begin{equation}
\label{eq:encoding}
 E_\A=\prod_{C\in\A}\prod_{j\in C\setminus\{a_C\}}\CX(j\to a_C).
\end{equation}
Different cached classes act on disjoint sets of data qubits, and the star-shaped CX gates within each class commute pairwise, so $E_\A^{-1}=E_\A$. For target $i\in C$, define the local compensation as
\begin{equation}
\label{eq:normalize}
K_i=\begin{cases}
 I,&C\notin\A,\\
 \CX(i\to a_C),&C\in\A,\ i\ne a_C,\\
 \prod_{j\in C\setminus\{i\}}\CX(j\to i),&C\in\A,\ i=a_C.
\end{cases}
\end{equation}
Represent $T_i=K_iE_\A$ as a matrix over $\GF$.
\begin{lemma}[Target Row and Column Isolation]\label{lem:rowcol}
Both row $i$ and column $i$ of $T_i$ are the unit vector $e_i$. Thus, the target bit in the temporary representation still corresponds to the original variable $x_i$, and flipping this target bit leaves the other temporary coordinates unchanged.
\end{lemma}
For a nonanchor target in a cached class, local compensation temporarily removes the target bit's encoding contribution to the occupancy anchor, so the local rotation preserves the original target variable's flip semantics. Once local compensation is undone, a legal flip updates the target bit and the class occupancy anchor together.

\subsection{Neighborhood Partitioning, Control Rank, and Parity Queries}

Order true-twin classes by increasing anchor. For target $i\in C$, its neighborhood consists of the remaining vertices $C\setminus\{i\}$ of its class and all neighboring true-twin classes. If its own class has remaining vertices and there is an external neighboring class, merge these vertices with the first external neighboring class in the prescribed class order. The other neighboring classes form separate disjoint blocks $S_{i1},\ldots,S_{ir_i}$.

Sort the blocks by size and lexicographically by vertices, fixing the order before cache selection. Because adjacency is complete within each true-twin class and between the distinct components of any merged block, each neighborhood block contains at most one occupied vertex on the independent-set feasible domain. Its parity therefore equals its occupancy.

Let $b_{ib}$ be the binary row mask of the original neighborhood block. Define the query row in temporary data coordinates as
\begin{equation}
\label{eq:pullback}
f_{ib}=b_{ib}T_i^{-1}.
\end{equation}
For temporary coordinates $y=T_ix$, we have $f_{ib}y=b_{ib}x$, so the transformed query row preserves the occupancy test for the original neighborhood block. Under the partitioning and target normalization rules above, all query rows are nonzero, their supports are pairwise disjoint, and none contains the target bit. The rows are therefore linearly independent, and the control rank is $r_i$. For each query row, the smallest-index qubit in its support is used as the pivot to compute the corresponding parity and construct the transformation $M_i$. The complete local macro is
\begin{equation}
\label{eq:macro-specific}
\widetilde U_i=K_i^\dagger M_i^\dagger\kappa_i(\beta)M_iK_i.
\end{equation}

Let $\delta_C$ denote the degree of class $C$ in the quotient graph. The partitioning rule gives
\begin{equation}
\label{eq:rank}
r_i=\begin{cases}
\delta_C,&|C|=1,\\
\max(1,\delta_C),&|C|>1.
\end{cases}
\end{equation}
The control rank is determined by the quotient-graph degree of the target class and whether it is a singleton. When the original neighborhood contains larger true-twin classes, the blockwise query can have a lower control rank than the number of vertices in that neighborhood, reducing the number of control conditions required for the local rotation.

\subsection{Coherent Query Synthesis and Automatic Restoration}
\label{sec:mechanism}
This subsection constructs open Gray--Walsh circuits and conditional phase query kernels and proves that both constructions satisfy a common restoration condition on the feasible subspace.

The local query kernel is selected piecewise by control rank. For $r_i=0$, a rotation without controls is used. For $1\le r_i\le3$, open Gray--Walsh synthesis uses $2^{r_i}$ parameterized rotations and retains the terminal phase frame. For $4\le r_i\le8$, a fixed coherent query kernel is used. Section B.4 of the Online Appendix gives its control-computation prefix and gate-level recipe.
For $r_i>8$, a scalable conditional phase kernel uses a deterministic control computation and its actual inverse. The construction selects two control wires as conditional workspace, computes an intermediate predicate on the remaining controls, and combines these operations into a complete query kernel. Structured control synthesis yields the gate-count bound for the high-rank kernel. Section B.5 of the Online Appendix gives the deterministic gate sequence and resource bound.

For a binary query row $f$, define the character $\chi_f(y)=(-1)^{fy}$ and let $\bar h_i(y)=\prod_b[1-(f_{ib}y\bmod2)]$ denote the compressed zero-control condition in encoded coordinates. Then
\begin{equation}
\label{eq:walsh}
 \bar h_i(y)=2^{-r_i}\sum_{J\subseteq[r_i]}\chi_{\oplus_{b\in J}f_b}(y).
\end{equation}
For $x\in\F$ and $y=T_i x$, we have $\bar h_i(T_i x)=h_i(x)$. Traverse the parity terms in a fixed Gray order, synthesize the rotations on the target qubit, and retain the terminal phase frame~\citep{welch}. Conjugating the remaining control transformation associated with closing the frame by a Hadamard on the target gives a CZ star operator. Its controls correspond to the original variables in the union of the neighborhood blocks. When the target bit is 1, all relevant neighborhood variables are zero, so this residual phase operator acts as the identity on the feasible output space. Section B.1 of the Online Appendix compares the resources of the open and closed implementations and gives the derivation.

For a conditional phase query kernel with control rank $r\ge4$, denote its unitary by $\kappa_r(\beta)$. Partition the $r$ control bits into sets $L$ and $R$, where $L$ contains two control bits and $R$ contains the remaining $r-2$. The two-control subkernel satisfies $B_2(\delta)=R_x(2\delta\NOR_L)D_B$, where $D_B$ is a parameter-independent diagonal involution that acts as the identity when $L=0$.
Pairing the computation prefix on the control register with its actual inverse and placing an exact CCX between them gives $A_r|t,z\rangle=|t\oplus q(z),z\rangle$, with all control bits restored to their original values. The complete query kernel is defined as

\begin{equation}
\label{eq:kernel-construction-main}
\kappa_r(\beta)=B_2(\beta/2)H_iA_rH_iB_2(-\beta/2)Z_i,
\end{equation}
where $B_2$ is implemented by a three-CX gate sequence, and the control-computation prefixes for ranks 4 through 8 use fixed recipes. This construction satisfies
\begin{equation}
\label{eq:kernel-involution}
 \kappa_r(\beta)=U_{\NOR,r}(\beta)D_r,\qquad
 D_r=Z_i^{1-q(z)},\quad D_r^2=I,\quad
 [D_r,U_{\NOR,r}(\beta)]=0,
\end{equation}
Here $q$ is independent of the target and parameters, and $q=\NOR_R$ when $L=0$. The residual operator $D_r$ acts as the identity on the admissible input domain specified in Eq.~\eqref{eq:care}.

The kernel has an analytical upper bound of $6r-12$ CX gates. In the experiments, prespecified gate-level recipes that satisfy the restoration contract are used for $4\le r\le8$, while the scalable kernel is used for $r>8$. The gate-level construction of the two-control subkernel $B_2$ and the control-computation operator $A_r$ is detailed in Sections B.3 and B.4 of the Online Appendix, respectively. The proofs of the residual operator properties and the half-angle pairing identity are given in Section B.6.

This gate-level construction also specifies the unitary action outside the admissible input domain and is hereafter referred to as the full-space completion kernel. Applying two completion kernels, each with parameter $\beta/2$, in sequence implements the canonical rotation exactly on the entire compressed target and control space for any real parameter value:
\begin{equation}
\label{eq:half-pair}
 \kappa_r(\beta/2)^2=U_{\NOR,r}(\beta),\qquad r\ge4.
\end{equation}

The analytical upper bound on the CX count of the corresponding circuit is $12r-24$. The matched full-space exact-kernel reference and our method use the same neighborhood partition, nonuniformly weighted cost function, and target execution order.

\begin{theorem}[Automatic Restoration]\label{thm:auto}
For any cache subset and the query kernels specified above for each control rank, the complete local mixer macros satisfy the conditions of Theorem~\ref{thm:composition}.
\end{theorem}
\begin{proof}[Proof Sketch]
$S_i=M_iK_iE_\A$ preserves target row and column isolation. After the preparation transformation $S_i$, each pivot bit on an original feasible input equals the occupancy flag of its neighborhood block. When $x_i=1$, all neighborhood blocks are empty, and the prepared state belongs to the admissible input domain defined by Eq.~\eqref{eq:care}. Actual decoding after an enabled flip satisfies
$K_i^{-1}M_i^{-1}(S_i x\oplus e_i)=E_\A(x\oplus e_i)$. When the control condition is inactive, the corresponding branch is unchanged. Linearly extending these equalities from the feasible computational basis states to the entire feasible subspace gives the local operator identity. Section A of the Online Appendix contains the complete proof and the related structural, cost, and amortization derivations.
\end{proof}

\subsection{Nonuniform Costs and Representation Maintenance Across Layers}
For $x\in\F$, the encoded bits $y$ on a cached class $C$ satisfy $y_{a_C}=\sum_{j\in C}x_j$ and $y_j=x_j$ for $j\ne a_C$, giving
\begin{equation}
\label{eq:cost}
 \sum_{j\in C}w_jx_j=w_{a_C}y_{a_C}+\sum_{j\ne a_C}(w_j-w_{a_C})y_j.
\end{equation}
For uncached classes, the cost layer temporarily computes the clique parity in the original coordinates and applies the inverse transformation after the phase operation to restore the data representation. With the same neighborhood partition, query synthesis rules, and other compilation settings, the caching schemes implement the same cost function. Their circuit resource differences reflect the cost of the persistent occupancy representation and its maintenance operations.

For a binary variable $y\in\{0,1\}$, the single-qubit diagonal unitary corresponding to the phase factor $e^{\ii\gamma cy}$ satisfies
\begin{equation}
\operatorname{diag}(1,e^{\ii\gamma c})=e^{\ii\gamma c/2}R_z(\gamma c).
\end{equation}
The compiler synthesizes each $R_z$ rotation from the actual cost coefficients in encoded coordinates and accumulates the corresponding global phase. With uniform weights, one parameterized rotation implements the cost phase of each cached class. With nonuniform weights, the weight differences for nonanchor variables are retained according to Eq.~\eqref{eq:cost}.

The intervening cost layers use independent parameters $\gamma_\ell$. When all cost coefficients in a layer are zero, the corresponding rotation gates can be omitted. When synthesis produces only a global phase, that phase is recorded in the circuit's global phase. Each cost layer implements the phase operations in the persistent encoding and preserves that representation for subsequent mixer operations, so the complete program requires encoding only at the beginning and decoding only at the end.

\subsection{Source-Level Cost Decomposition for Cache Selection and Amortization over Multiple Sweeps}
Let $z_C\in\{0,1\}$ indicate whether class $C$ is cached. In the following analysis, the choice of anchors, neighborhood partition, parity factor order, Gray traversal order, query kernel recipe, pivot selection rule, and initial layout are all fixed, and each source-level CX gate is assigned a prespecified nonnegative weight. The source-level cost includes both encoding and decoding boundaries, all mixer operations, and the intervening cost layers.
\begin{theorem}[Source-Level Cost Separability under a Fixed Recipe]\label{thm:separable}
The objective $J$ above satisfies
\begin{equation}
\label{eq:affine}
J(\A)=J(\varnothing)+\sum_{C\in\A}\bigl[J(\{C\})-J(\varnothing)\bigr].
\end{equation}
Consequently, all classes with strictly negative marginal cost form an optimal cache subset for this fixed source-level objective. Classes with positive marginal cost are left uncached, and those with zero marginal cost are left uncached by the prescribed rule. The result also holds when CX gates are charged by the distance between their endpoints under a fixed layout.
\end{theorem}
\begin{proof}[Proof Sketch]
Caching a module changes only its contributions to boundary encoding, target compensation, and parity factors, and the contributions of distinct modules do not overlap. The lowest-index support position for an external module is always the anchor, and the lowest-index support position for the target module also remains fixed. The caching decision therefore leaves pivot selection, the query mapping, and instructions for other modules unchanged. Hence the endpoint multiplicity vector of source-level CX gates is an affine function of the cache indicators $\{z_C\}_{C\in\C}$.
Summing with the fixed per-gate weights yields Eq.~\eqref{eq:affine}.
\end{proof}

Exact source-level cost separability reduces cache selection to an independent marginal cost test for each true-twin class. With a fixed circuit recipe and fixed per-gate weights, evaluating the no-cache scheme and each single-class cache scheme determines the source-level optimal cache subset. Complete compilation then synthesizes the source-level optimum, empty-cache, and full-cache candidates under a common backend configuration and selects the output based on complete-circuit resources, accounting for the actual effects of cross-macro gate cancellation and topology routing.

\paragraph{Amortization over Multiple Sweeps and Weighted Source-Level Cost.}
Consider amortization over multiple sweeps with unit CX weights. Let $B(\A)=\sum_{C\in\A}(|C|-1)$ denote the source-level CX count at one encoding or decoding boundary, and let $F(\A)$ and $H(\A)$ denote the source-level CX counts of one complete mixer sweep and one cost layer, respectively. If all sweeps use the same local mixer macro recipes and all cost layers use the same weighting pattern, then
\begin{equation}
\label{eq:amortization}
J_p(\A)=2B(\A)+pF(\A)+(p-1)H(\A).
\end{equation}
Let $\Delta_F=F(\varnothing)-F(\A)$ and $\Delta_H=H(\varnothing)-H(\A)$. A strict gain is equivalent to
\begin{equation}
\label{eq:breakeven}
p\Delta_F+(p-1)\Delta_H>2B(\A).
\end{equation}
When the sweeps use different mixer macro recipes or the cost layers use different weight configurations, compute the respective source-level resources and sum them. For general fixed endpoint weights $\rho$, define
\begin{equation}
B_\rho=\sum_{C\in\A}\sum_{j\in C\setminus\{a_C\}}\rho(j,a_C).
\end{equation}
Evaluating $F$ and $H$ with the same endpoint weights and replacing $B$ by $B_\rho$ extends the amortization criterion to weighted source-level costs. The selected cache subset achieves a strict source-level benefit under the fixed recipe when the cumulative weighted CX savings from the sweeps and cost layers exceed the cost of encoding and decoding at both boundaries.

With a fixed upper bound $\Delta$ on the quotient-graph degree, uniform clique occupancy costs, and a bounded-rank query kernel library, the source-level circuit that caches all nontrivial classes uses $O(pn(\Delta+1)+n)$ CX gates. Each cost layer uses at most one parameterized rotation per class.

\subsection{Complete Circuit Selection and Restoration of the Original Wire Order}

After determining the source-level optimal cache subset by Theorem~\ref{thm:separable}, the compiler synthesizes three classes of candidates: the source-level optimum, empty cache, and full cache. Duplicate cache schemes are merged. Each candidate undergoes complete circuit synthesis under the same backend configuration. The candidates are sorted lexicographically by CX count, number of parameterized rotations, and circuit depth, with ties in resources broken by a fixed candidate order to select the final output circuit. All resource metrics are computed from that output circuit.

For a given linear or two-dimensional grid topology, the compiler generates source-level candidates using endpoint distances under the fixed initial layout and performs physical routing on the candidate circuits. Swaps along device coupling edges then restore the original wire order of the logical qubits. The CX count and depth of restoration are included in the final resource counts, and the global phase of the complete circuit is preserved.

\section{Experimental Setup and Performance Evaluation}
\label{sec:evaluation}
This section uses fixed reference implementations, independent confirmation, and matched ablations to evaluate the CX savings of complete circuits and their sources. It analyzes the effects of true-twin class structure, the number of mixer sweeps, cost weights, and physical topology. Public-graph and scaling experiments examine compilation performance on different graph structures and classical compilation overhead, respectively.

\subsection{Benchmark Programs, Baselines, and Common Backend}
\paragraph{Tasks and metrics.}
The main evaluation set covers $n=32,64,128$ and quotient-graph degrees $2,4,6,8$. It includes a pattern in which all true-twin classes have two vertices and a mixed pattern with one-, two-, and four-vertex classes. Graphs are deduplicated by their labeled vertices and edge sets, and duplicate programs are removed based on the input graph, number of complete mixer sweeps, weight configuration, and target execution order. This yields a main evaluation set of 62 distinct labeled graphs and 134 ordered programs.

The independent confirmation set is constructed separately after the compilation rules are fixed and covers $n=48,96,192$ and quotient-graph degrees $3,5,7$. Exact isomorphism checking on quotient graphs that preserve module-size attributes yields 36 nonisomorphic graphs and 72 ordered programs. Both sets use two complete mixer sweeps, uniform weights, and one intervening cost layer, with independent parameters $\beta_0,\beta_1,\gamma_0$. Within each program, both sweeps use the same target execution order, and each graph is evaluated with an ascending order and a fixed pseudorandom order. Section C.1 of the Online Appendix gives the generation and deduplication rules and random seed settings.

In addition to the main and independent confirmation sets, the experiments include public benchmark graphs, fixed small instances for correctness checks, and scaling instances. The main and confirmation sets are constructed from connected regular quotient graphs by replacing vertices with true-twin cliques to control module size and quotient-graph degree. Public graphs are used to evaluate other graph structures, fixed instances are used to check correctness and mechanisms, and scaling instances are used to measure classical compilation overhead.

The primary metric is the CX count of the complete circuit. We also report circuit depth, parameterized $R_z$ count, single-qubit gate count, and classical compilation time. The reference CX envelope for each program is the lowest CX count among the relevant fixed reference implementations, and the paired reduction is $1-C_{\rm ours}/C_{\rm ref}$. Multi-resource comparisons use the complete resource vector of one actual circuit. Reference lower envelopes attained separately for each metric are listed separately.

\paragraph{Fixed baselines.}
The six fixed reference implementations fall into three groups. The first consists of two structured/Vale implementations that combine the related structured constructions of \citet{compact25} with Vale controlled-rotation decomposition~\citep{vale}, using the original cost and the clique cost, respectively. The second comprises open and closed CPG--Walsh synthesis~\citep{bergholm05,mottonen,welch}. The third consists of two matched full-space exact kernel implementations that share the automatic partitioning frontend used by our method and use the original cost and clique cost, respectively. The first two groups cover different conditional-rotation synthesis methods. Sharing the automatic partitioning frontend allows the third group to control for differences in neighborhood partitioning in the resource comparison.

The structured/Vale implementations use the smallest pivot and actual uncomputation. For control ranks $r\le8$, the matched exact kernels use a prespecified library of full-space exact kernels; for $r>8$, they use the exact half-angle pairing in Eq.~\eqref{eq:half-pair}. All implementations use the same number of data qubits, target execution order, and layer parameters, and preserve the same cost phase action on the feasible subspace. Implementations with different cost representations are measured separately to analyze the effect of the cost representation on circuit resources.

Additional comparisons include direct decomposition of the canonical NOR-controlled operation and the fixed-order and tree configurations of the QuCLEAR implementation of \citet{quclear}. Each implementation retains the required terminal restoration and phase operations under its complete gate-level recipe. The six main references cover all programs in the main and independent confirmation sets, and the three main references for the public-graph experiment cover all 24 programs. Additional references are compared on the common set of programs for which they complete successfully under matching experimental specifications. Section C.2 of the Online Appendix gives the specific recipes for the fixed reference implementations.

\paragraph{Native synthesis and physical routing.}
All implementations use the same Qiskit 2.1.2 backend and native gate set $\{R_z,\mathrm{SX},X,\CX\}$. Following translation into the native gate set, the same bounded number of rounds of single-qubit gate simplification and adjacent CX/X inverse cancellation is applied, with the symbolic global phase retained.

The physical routing experiments use three fixed seeds for each selected program and topology. Both the linear topology and the rectangular two-dimensional grid use the identity initial layout and a fixed SABRE configuration. The post-routing resource counts include terminal wire-order restoration.

\subsection{Complete-Circuit Resources and Independent Confirmation}
Table~\ref{tab:main} summarizes the native circuit resources of the six fixed reference implementations, the per-program reference CX envelope, and our complete output. Across the 134 programs in the main evaluation set, the total CX count falls from 611,250 to 473,016, a reduction of 22.61\%. Across the 72 programs in the independent confirmation set, it falls from 442,568 to 354,650, a reduction of 19.87\%. For every program in both sets, the output CX count is below the corresponding reference envelope, and the total circuit depth in the independent confirmation set is 344,505. These results show that structure-aware compilation yields complete-circuit CX savings for ordered programs in both sets, which differ in graph size and quotient-graph degree.

\begin{table}[H]
\centering
\caption{Complete native circuit resources for the main evaluation and independent confirmation sets}
\label{tab:main}

\begin{threeparttable}
\scriptsize
\setlength{\tabcolsep}{4pt}
\renewcommand{\arraystretch}{1.05}

\begin{tabular*}{\textwidth}{@{\extracolsep{\fill}}p{.26\linewidth}rrrrr@{}}
\toprule
Implementation & Main CX$\downarrow$ & Confirm. CX$\downarrow$ & Confirm. depth$\downarrow$ & Confirm. 1Q$\downarrow$ & \shortstack{Confirm.\\param. rotations$\downarrow$}\\
\midrule
Structured/Vale / original cost & 1,039,536 & 801,648 & 990,223 & 1,622,326 & 72,576\\
Structured/Vale / clique cost & 1,034,528 & 797,490 & 990,128 & 1,618,294 & \textbf{68,544}\\
Open CPG & 2,927,610 & 1,487,440 & 1,270,996 & 1,003,968 & 907,200\\
Closed CPG & 2,998,668 & 1,545,744 & 1,327,273 & 1,003,968 & 907,200\\
Matched exact kernel / original cost & 620,368 & 446,832 & 420,307 & 667,404 & 180,096\\
Matched exact kernel / clique cost & 615,360 & 442,674 & 420,220 & \textbf{663,372} & 176,064\\
\midrule
 Reference CX envelope & 611,250 & 442,568 & --- & --- & ---\\
Our complete output & \textbf{473,016} & \textbf{354,650} & \textbf{344,505} & 690,424 & 133,056\\
\bottomrule
\end{tabular*}

\begin{tablenotes}[flushleft]
\scriptsize
\item The reference CX envelope takes the lowest CX count across the six fixed reference implementations for each program. It need not correspond to a single fixed reference circuit, so the other resource columns are left blank. 1Q denotes the single-qubit gate count. All resource metrics for our method come from the same final output circuit, and bold indicates the lowest measured value in the corresponding column.
\end{tablenotes}
\end{threeparttable}
\end{table}

Table~\ref{tab:supplement-comparison} provides complete CX comparisons on the common set of programs covered by the additional references and under Qiskit optimization level 1. On the 12 programs with 32 qubits for which all nine reference implementations complete, our method reduces the total CX count from 5,780 to 4,478, a reduction of 22.53\%. Under Qiskit optimization level 1, the total CX counts decrease by 22.60\% and 19.83\% for the 134 main evaluation programs and the 72 independent confirmation programs, respectively, with a CX reduction for every paired program. The similar CX reductions under the two transpiler optimization configurations indicate that the CX advantage of our method remains stable under this change in configuration. Section D.1 of the Online Appendix describes reference coverage and the specific configurations.

\begin{table}[H]
\centering
\caption{Complete CX comparisons on the common set of additional references and under Qiskit optimization level 1}
\label{tab:supplement-comparison}

\begin{threeparttable}
\scriptsize
\setlength{\tabcolsep}{4pt}
\renewcommand{\arraystretch}{1.05}

\begin{tabular*}{\textwidth}{@{\extracolsep{\fill}}lrrrr@{}}
\toprule
Paired set / configuration & Programs & Reference CX & Our CX & Total CX reduction\\
\midrule
Common set of nine references (32 qubits) & 12 & 5,780 & 4,478 & 22.53\%\\
Level 1, main evaluation set & 134 & 611,134 & 473,016 & 22.60\%\\
Level 1, independent confirmation set & 72 & 442,392 & 354,650 & 19.83\%\\
\bottomrule
\end{tabular*}

\begin{tablenotes}[flushleft]
\scriptsize
\item The first row corresponds to the 12 programs with 32 qubits for which all nine references complete. The other two rows use Qiskit optimization level 1 and the per-program reference CX envelopes for their respective program sets.
\end{tablenotes}
\end{threeparttable}
\end{table}

\begin{figure}[H]
\centering
\Figure{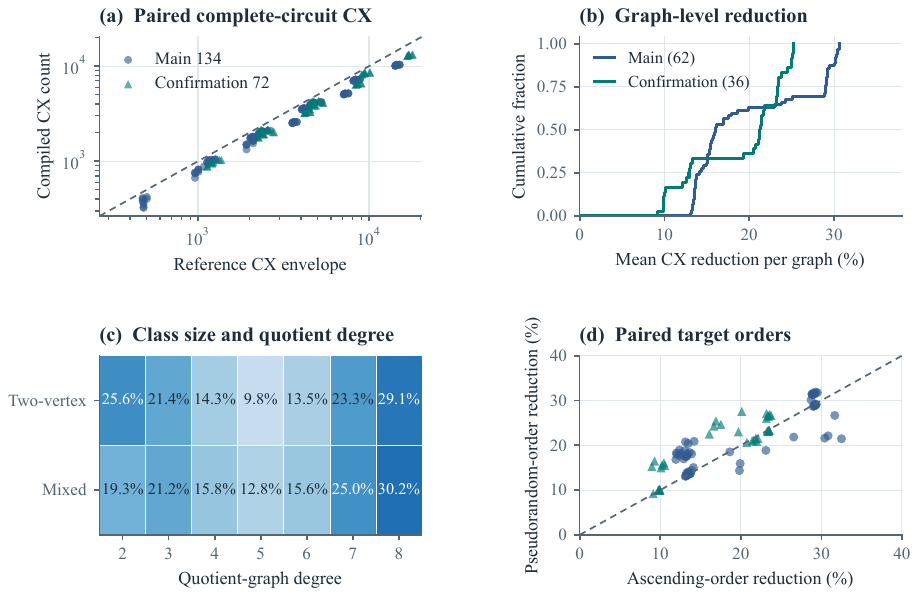}
\caption{CX counts and distributions of paired savings in the main evaluation and independent confirmation sets}
\label{fig:mainnew}
\FigureNote{(a) Per-program reference CX envelope versus the CX count of our circuit. The diagonal denotes equality. (b) Empirical cumulative distributions of the mean per-program CX reduction within each graph, covering 62 labeled graphs and 36 nonisomorphic graphs, respectively. (c) Mean per-graph reductions aggregated by quotient-graph degree and module pattern. The two sets cover even and odd degrees, respectively. (d) Paired results for ascending and fixed pseudorandom target orders on the same graph.}
\end{figure}

Figure~\ref{fig:mainnew} shows per-program CX savings and their distribution across graph instances, quotient-graph degrees, and target execution orders. The main evaluation set covers even quotient-graph degrees, and the independent confirmation set covers odd degrees. The per-program CX reductions in the confirmation set range from 8.96\% to 27.57\%, with positive CX savings across its graph sizes and odd-degree settings. For graph-level statistics, the CX reductions for different target execution orders are first averaged within each graph, and the results are then aggregated with equal weight per graph. For the 62 graphs in the main evaluation set, the mean, median, and interquartile interval of the graph-level reductions are 20.08\%, 16.01\%, and [14.21\%, 29.02\%], respectively. The corresponding statistics for the 36 graphs in the independent confirmation set are 18.92\%, 21.28\%, and [12.92\%, 23.37\%]. Section D.2 of the Online Appendix gives the per-program resource records for the independent confirmation set.

\subsection{Ablation of Compilation Mechanisms and Effects of Graph Structure}

The CX savings of complete compilation arise from both local conditional query synthesis and occupancy reuse across operations. Table~\ref{tab:attribution} compares the fixed reference CX envelope, the circuit using the same query synthesis mechanism with no persistent caching, and our complete output to evaluate the overall CX savings and the incremental contribution of persistent encoding. The net CX reductions from caching are 4.10\%, 9.13\%, 3.64\%, and 0.47\% for the main evaluation set, independent confirmation set, sweep and weight evaluation set, and public-graph evaluation set, respectively.

\begin{table}[H]
\centering
\caption{CX savings of complete compilation and the additional contribution of persistent caching}
\label{tab:attribution}

\begin{threeparttable}
\scriptsize
\setlength{\tabcolsep}{4pt}
\renewcommand{\arraystretch}{1.05}

\begin{tabular*}{\textwidth}{@{\extracolsep{\fill}}>{\raggedright\arraybackslash}p{.27\linewidth}rrrrr@{}}
\toprule
Evaluation set & \shortstack{Ref. CX\\envelope} & \shortstack{No-cache\\CX} & \shortstack{Complete\\output CX} & \shortstack{Cache\\reduction} & \shortstack{Share of\\savings}\\
\midrule
Main evaluation (134 programs) & 611,250 & 493,228 & 473,016 & 4.10\% & 14.62\%\\
Independent confirmation (72 programs) & 442,568 & 390,280 & 354,650 & 9.13\% & 40.53\%\\
Sweeps and weights (120 programs) & 972,132 & 856,548 & 825,400 & 3.64\% & 21.23\%\\
Public graphs (24 programs) & 205,334 & 167,404 & 166,620 & 0.47\% & 2.03\%\\
\bottomrule
\end{tabular*}

\begin{tablenotes}[flushleft]
\scriptsize
\item The net cache reduction is the difference in CX count between the circuit without persistent caching and the complete output, divided by the CX count of the circuit without persistent caching. The cache share of savings is the additional CX savings from persistent caching divided by the total CX savings between the reference envelope and the complete output.
\end{tablenotes}
\end{threeparttable}
\end{table}

\begin{figure}[H]
\centering
\Figure{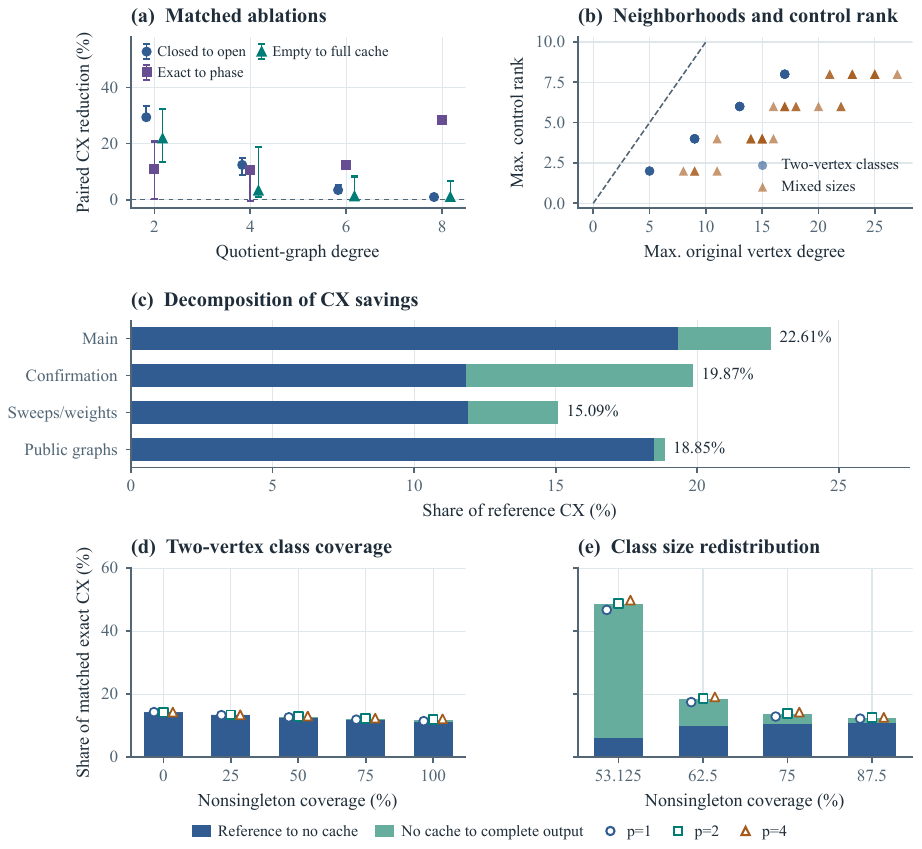}
\caption{CX savings from local queries, persistent caching, and true-twin class structure}
\label{fig:mechanismnew}
\FigureNote{(a) Three matched ablations for the 134 main evaluation programs. Each reduction uses the CX count of the first scheme in the corresponding comparison as the denominator. Markers denote medians, and bars denote observed ranges. The caching comparison uses the empty and full caches. (b) Maximum vertex degree and maximum control rank for the 62 labeled graphs. The dashed line denotes equality. (c) Decomposition of total savings using the fixed reference CX envelope for each set as the denominator. (d,e) The coverage of two-vertex classes and the allocation of class sizes on a fixed quotient graph are varied, respectively. Bar heights denote mean reductions for each structural condition, using each program's matched exact kernel CX count as the denominator. Marker shapes distinguish sweep counts, and markers denote the mean CX reduction of the complete output relative to the matched exact kernel across a total of 72 controlled programs.}
\end{figure}

Figure~\ref{fig:mechanismnew} further shows the ablations of compilation mechanisms and the effect of true-twin class structure on CX savings. In the main evaluation set, the closed and open CPG--Walsh circuits have total CX counts of 2,998,668 and 2,927,610, respectively. Leaving the terminal frame open reduces the CX count by 2.37\%. With the same automatic partitioning frontend and clique cost, the matched full-space exact kernel reference and the phase query circuit without persistent caching have total CX counts of 615,360 and 493,228, respectively, a reduction of 19.85\% for the latter. Adding persistent encoding and performing candidate selection further reduces the total CX count to 473,016, a reduction of 4.10\% relative to the no-cache scheme.

To examine the effects of true-twin class coverage and the distribution of class sizes, we set up two controlled experiments on graph structure with 32 vertices, quotient-graph degree 4, and uniform weights. Coverage is defined as $\mathrm{coverage}=\sum_{C\in\C,\,|C|>1}|C|/n$. The first experiment varies the proportions of singleton and two-vertex classes to achieve coverages of 0\%, 25\%, 50\%, 75\%, and 100\%. The second fixes the quotient graph at 16 vertices and varies the allocation of sizes across true-twin classes. The 12 graphs are each combined with two target execution orders and three sweep counts $p=1,2,4$, yielding 72 programs.

Figure~\ref{fig:mechanismnew}(d,e) gives the results of the two controlled experiments on graph structure. The additional CX savings from persistent caching are measured as $\bigl(C_{\mathrm{nocache}}-C_{\mathrm{out}}\bigr)/C_{\mathrm{matched}}$. At zero coverage, phase query synthesis reduces the CX count by 14.29\% relative to the matched exact kernel. In the series that combines singleton and two-vertex classes, the additional savings from caching reach 1.19\%. When the additional vertices are concentrated in one 17-vertex true-twin class, the additional savings from caching range from 39.16\% to 45.01\% across the six corresponding programs. At fixed vertex count and quotient-graph degree, the controlled results show that the distribution of true-twin class sizes has an important effect on the benefits of persistent caching. Concentrating vertices in a larger true-twin class can yield more pronounced savings from reusing occupancy information across operations.

\subsection{Cache Amortization over Multiple Mixer Sweeps and under Nonuniform Weights}

To study the amortization benefits of persistent occupancy encoding over multiple mixer sweeps and the effect of nonuniform weights on circuit resources, we use 12 fixed generated inputs with four sweep counts $p=1,2,4,8$ and three cost patterns: uniform, periodic nonuniform, and two-level weights. For $p=1$, there is no intervening cost layer and the different weight patterns yield identical circuits. Deduplication gives 120 ordered programs, with independent cost parameters across layers. The experiments consider both the source-level amortization condition in Eq.~\eqref{eq:breakeven} and the actual resources of the complete backend circuits.

\begin{figure}[H]
\centering
\Figure{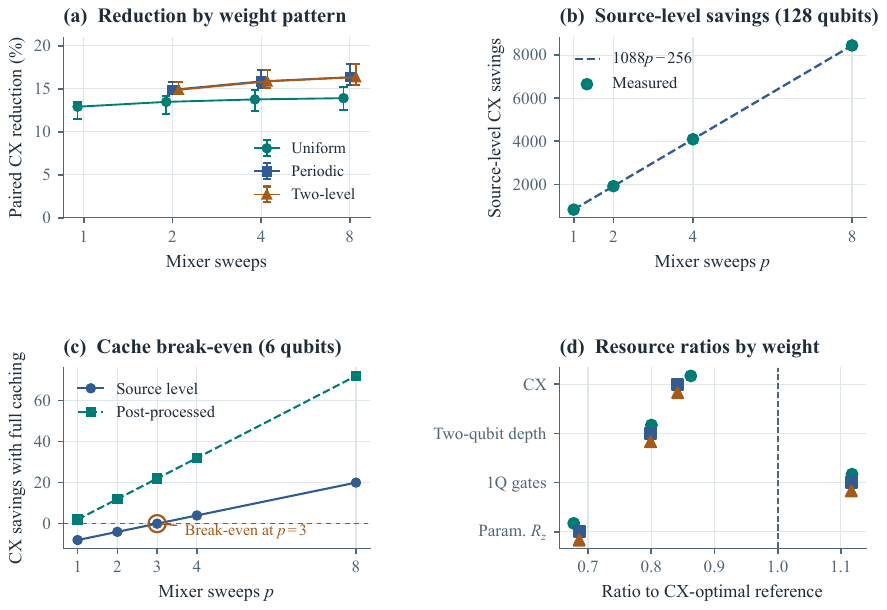}
\caption{Cache amortization for different mixer sweep counts and weight patterns}
\label{fig:amortization}
\FigureNote{(a) Mean savings and observed ranges for the 12 fixed inputs. Points are shifted horizontally to distinguish weight patterns at the same sweep count. Nonuniform weights are compared only for $p\ge2$. (b) Source-level CX savings and their affine relation for the 128-vertex mixed-module instance. (c) Source-level and post-processing CX savings in the fixed six-vertex example, with the source-level costs tying at $p=3$. (d) Statistics are computed only for the shared conditions with $p\ge2$. The reference resource vector for each program comes from its corresponding CX-optimal reference circuit.}
\end{figure}

Figure~\ref{fig:amortization} shows the circuit savings and source-level amortization results for different sweep counts and weight patterns. Relative to the per-program CX envelope of three fixed reference implementations, the total CX count for the 120 programs falls from 972,132 to 825,400, a decrease of 15.09\%, with a CX reduction for every program. The total complete-circuit depth decreases by 22.12\%. For sweep counts $p=1,2,4,8$, the aggregate CX reductions are 12.96\%, 14.35\%, 15.05\%, and 15.39\%, respectively. The CX reduction increases with the sweep count in this experiment, consistent with the amortization of the encoding boundary cost over multiple mixer sweeps.

For the 36 graph structure--weight configuration combinations, the difference in source-level caching cost under the fixed recipe follows the affine relation in Eq.~\eqref{eq:amortization} for every tested sweep count. For a fixed mixed-module input with 128 vertices and quotient-graph degree 4, the source-level CX savings under uniform weights satisfy $J_p(\varnothing)-J_p(\C)=1088p-256$. The coefficient transformation for cost layers in programs with nonuniform weights is included in the source-level amortization model, while complete-circuit resources are measured from the backend synthesis outputs. Figure~\ref{fig:amortization}(c) further presents a six-vertex example whose quotient graph is a three-vertex path $P_3$ and whose true-twin classes all contain two vertices. The source-level CX savings of full caching relative to no caching are $4p-12$. At $p=3$, both schemes use 90 source-level CX gates, while the post-processed no-cache and full-cache circuits contain 90 and 68 CX gates, respectively. Full caching therefore saves 22 actual CX gates when the source-level costs tie, showing that backend gate optimization can change the resource difference predicted by the source-level cost model.

\subsection{Public-Graph Instances and Physical Topology Routing}
\label{sec:public-routing}

The public-graph evaluation set comprises 12 graphs from the CMU COLOR graph coloring benchmark~\citep{cmucolor}, including graphs from the Mycielski construction~\citep{mycielski1955} and Queen, character association, and sports schedule graphs from the Stanford GraphBase~\citep{knuth1993graphbase}. We retain each original graph's vertices and undirected edges, interpret each edge $\{u,v\}$ as an independent-set constraint $x_ux_v=0$, and construct a constraint-preserving mixer program according to Section~\ref{sec:spec}~\citep{hadfield}. Each graph uses uniform weights and two complete mixer sweeps, with ascending and fixed pseudorandom execution orders considered separately, yielding 24 ordered programs.

For each public graph instance, Table~\ref{tab:public-attribution} lists the number of nontrivial true-twin classes, the CX counts of the no-cache and complete circuits under the two orders, and the net reduction from caching. Four graphs have nontrivial true-twin classes identifiable by closed-neighborhood equivalence. Persistent caching yields additional CX reductions of 1.64\%, 6.06\%, and 1.63\% for David, Huck, and Jean, respectively. For the other nine public graphs, the complete output and no-cache scheme have the same CX count. This control uses the same query synthesis and cost implementation to evaluate the incremental contribution of persistent caching. Section D.3 of the Online Appendix provides the complete per-program resources.

\begin{table}[H]
\centering
\caption{Additional CX savings from persistent caching on public graph instances}
\label{tab:public-attribution}

\begin{threeparttable}
\scriptsize
\setlength{\tabcolsep}{4pt}
\renewcommand{\arraystretch}{1.05}

\begin{tabular*}{\textwidth}{@{\extracolsep{\fill}}lrrrr@{}}
\toprule
Graph & Nontrivial true-twin classes & No-cache CX (A/S) & Complete CX (A/S) & Net cache reduction\\
\midrule
david.col & 9 & 6,904/6,874 & 6,780/6,772 & 1.64\%\\
games120.col & 1 & 12,326/12,330 & 12,326/12,330 & 0.00\%\\
huck.col & 11 & 3,660/3,634 & 3,432/3,420 & 6.06\%\\
jean.col & 7 & 3,560/3,568 & 3,500/3,512 & 1.63\%\\
myciel3.col & 0 & 226/226 & 226/226 & 0.00\%\\
myciel4.col & 0 & 1,152/1,152 & 1,152/1,152 & 0.00\%\\
myciel5.col & 0 & 4,536/4,536 & 4,536/4,536 & 0.00\%\\
myciel6.col & 0 & 15,840/15,840 & 15,840/15,840 & 0.00\%\\
queen5\_5.col & 0 & 3,240/3,240 & 3,240/3,240 & 0.00\%\\
queen6\_6.col & 0 & 6,096/6,096 & 6,096/6,096 & 0.00\%\\
queen7\_7.col & 0 & 10,248/10,248 & 10,248/10,248 & 0.00\%\\
queen8\_8.col & 0 & 15,936/15,936 & 15,936/15,936 & 0.00\%\\
\bottomrule
\end{tabular*}

\begin{tablenotes}[flushleft]
\scriptsize
\item Nontrivial classes are identified by closed-neighborhood equivalence. A denotes ascending order and S denotes the fixed pseudorandom order. The no-cache and complete outputs use the same query and cost implementations. The net reduction is computed from the CX counts summed across the two orders.
\end{tablenotes}
\end{threeparttable}
\end{table}

\begin{figure}[H]
\centering
\Figure{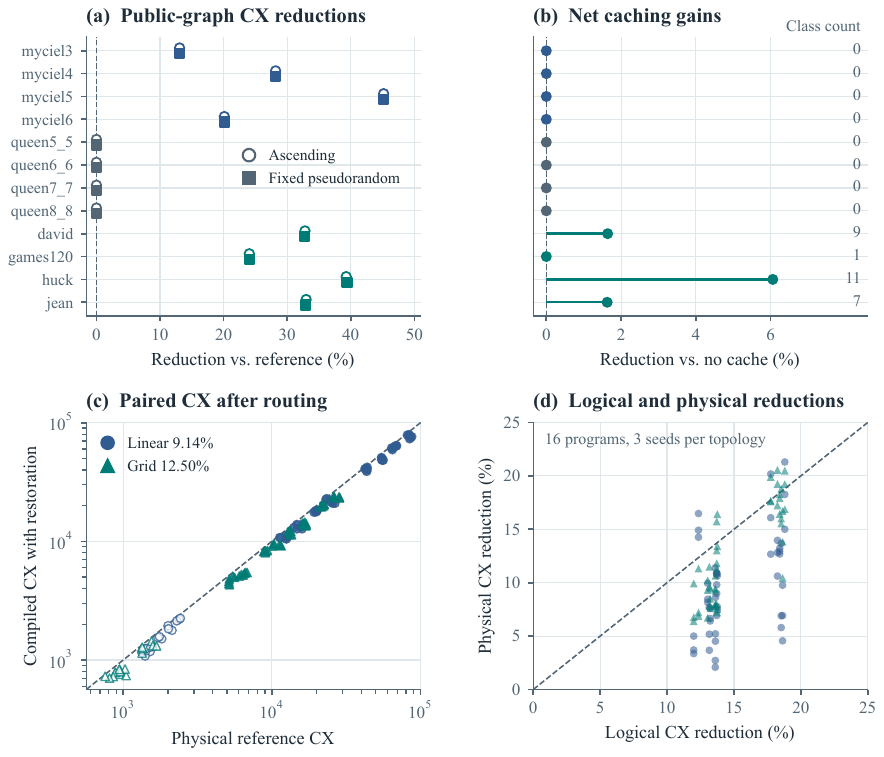}
\caption{CX savings in the public-graph and physical routing experiments}
\label{fig:public-routing}
\FigureNote{(a,b) The 24 programs from the 12 public graphs. (a) Open circles denote ascending order, and filled squares denote the fixed pseudorandom order. Reductions are relative to the lowest CX count per program across three fixed reference implementations. (b) Net cache reduction pooled across both orders, using the no-cache CX count as the denominator. The right side shows the number of nontrivial true-twin classes. (c,d) The physical routing set, with 66 program--seed pairs each for the linear and grid topologies. (c) Filled markers denote the main evaluation subset, and open markers denote fixed graphs used for mechanism checks. The physical reference is the minimum CX count across six references. Our counts include terminal restoration. (d) Logical reductions and per-run physical reductions for the 16 main evaluation programs.}
\end{figure}

Figure~\ref{fig:public-routing}(a,b) shows CX savings on the public graphs. Relative to the per-program reference CX envelope of open CPG and the two matched exact kernels, the total CX count across the 24 programs decreases from 205,334 to 166,620, a reduction of 18.85\%. Sixteen programs have lower CX counts than the reference and the remaining eight tie it. Including the additional references that completed leaves the per-program reference CX envelope unchanged.

The total CX reductions for the Mycielski graphs and the character association and sports schedule graphs are 27.40\% and 29.98\%, respectively. The four Queen instances use the same high-rank query implementation as the reference and tie its CX counts. For the public-graph set, local query synthesis is the main source of the overall 18.85\% CX reduction. Persistent occupancy encoding saves a further 0.47\% of the total CX count with the same query synthesis mechanism. Together with the David, Huck, and Jean results, this shows that the additional caching savings are concentrated in some graphs with reusable true-twin structure.

The physical routing experiment includes 16 programs from the main evaluation set and six programs formed from three fixed graphs used for mechanism checks, each with two orders, for a total of 22 programs. Each program is run with three fixed seeds on each of the two topologies, yielding 66 program--seed pairs per topology. The six main references complete under all conditions. The reference CX envelope takes the minimum for each matched program, topology, and seed. Table I-6 in the Online Appendix separately records the CX costs of the routing input, final physical output, and terminal wire-order restoration. Section C.3 of the Online Appendix gives the routing and restoration protocol.

Figure~\ref{fig:public-routing}(c,d) gives the CX savings after physical routing. Including terminal wire-order restoration, the total CX count on the linear topology decreases from the reference count of 1,960,198 to 1,781,104, a reduction of 9.14\%. On the two-dimensional grid, it decreases from 684,282 to 598,736, a reduction of 12.50\%. CX savings are obtained in all 66 pairs on each topology. For each program, the paired reductions are averaged across the three routing seeds, and the savings distribution is then computed over the 22 programs. The median reductions on the linear and two-dimensional grid topologies are 9.54\% and 12.76\%, with interquartile intervals [7.09\%, 12.83\%] and [8.86\%, 16.46\%], respectively.

\subsection{Multi-Resource Trade-Offs and Classical Compilation Overhead}

\begin{figure}[H]
\centering
\Figure{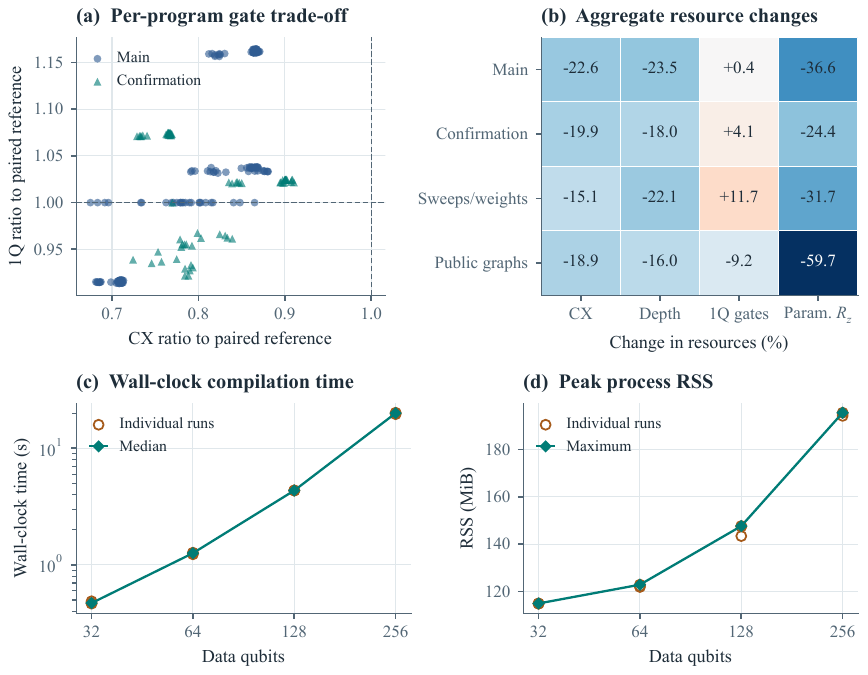}
\caption{Gate-resource trade-offs and classical compilation overhead of complete circuits}
\label{fig:resource-cost}
\FigureNote{(a) Ratios of per-program CX and single-qubit gate counts to the paired CX-optimal reference in the main evaluation and independent confirmation sets. The dashed line is at 1. (b) Resource changes computed after aggregating each resource across our method and the same reference circuits by set. Negative values denote reductions. (c,d) Complete native compilation overhead in the scaling experiment. Measurements from three independent processes are shown for each size. (c) The solid line denotes the median wall-clock time. (d) The solid line denotes the maximum peak RSS.}
\end{figure}

Building on the CX evaluation, this subsection examines multi-resource trade-offs and classical overhead under CX-first compilation. Figure~\ref{fig:resource-cost}(a,b) shows the per-program trade-off between CX and single-qubit gate counts and the changes in resource totals for each set, respectively. The comparison selects the actual reference circuit with the lowest CX count per program and uses its complete resource vector to evaluate the changes in depth, single-qubit gate count, and parameterized rotation count that accompany CX-first optimization. Under this comparison, the independent confirmation set shows reductions of 19.87\% in total CX count, 18.04\% in total depth, and 24.43\% in parameterized rotation count, with a 4.10\% increase in single-qubit gate count.

For each evaluation set, we also take the minimum of each resource metric separately, program by program, across the fixed reference implementations for that set to obtain an independent reference lower envelope for that metric. Under this comparison, total depth and two-qubit depth in the independent confirmation set decrease by 18.01\% and 12.82\%, respectively, while single-qubit gate count and parameterized rotation count increase by 7.12\% and 94.12\%, respectively. The latter two metrics increase by 17.93\% and 86.51\% in the sweep and weight evaluation set and decrease by 6.05\% and 34.28\% in the public-graph evaluation set. The CX-first paired reference and the independent per-resource lower envelopes use different comparison criteria. The resource vectors of our complete output circuits are the same in both comparisons.

We select candidates lexicographically with CX count as the primary metric and also report the depth, single-qubit gate count, and parameterized rotation count of the selected circuit. In the independent confirmation set, the CX-first outputs reduce both total CX count and circuit depth but use 690,424 single-qubit gates, 27,052 more than the fixed reference implementation with the lowest total single-qubit gate count among the six.

For classical compilation overhead, Table~\ref{tab:full-entry} reports the number of candidates actually synthesized, wall-clock time, CPU time, and peak resident memory by problem size. Figure~\ref{fig:resource-cost}(c,d) shows how compilation time and memory overhead vary with size. In the 256-vertex instance with 128 true-twin classes, source-level selection and full caching yield the same subset. After deduplication, only the empty-cache and full-cache candidates are synthesized. The median wall-clock time for complete compilation is 20.04 seconds, and the largest peak RSS across the three independent Linux processes is 195.5 MiB.

\begin{table}[H]
\centering
\caption{Time and memory overhead of complete native compilation}
\label{tab:full-entry}

\begin{threeparttable}
\scriptsize
\setlength{\tabcolsep}{4pt}
\renewcommand{\arraystretch}{1.05}

\begin{tabular*}{\textwidth}{@{\extracolsep{\fill}}rrrrrr@{}}
\toprule
$n$ & True-twin classes & Synth. candidates & Median wall (s) & Median CPU (s) & Max RSS (MiB)\\
\midrule
32 & 16 & 2 & 0.47 & 0.47 & 115.0\\
64 & 32 & 2 & 1.26 & 1.26 & 122.9\\
128 & 64 & 2 & 4.35 & 4.35 & 147.6\\
256 & 128 & 2 & 20.04 & 20.03 & 195.5\\
\bottomrule
\end{tabular*}

\begin{tablenotes}[flushleft]
\scriptsize
\item The experiment uses generated instances with quotient-graph degree 4 and two vertices per true-twin class. Each size is run in three single-threaded Linux processes (Intel Xeon Gold 6448Y, Python 3.12.3), with up to 12 tasks running concurrently. Wall-clock timing starts at structure discovery after importing the runtime and ends when the final native circuit is selected. The peak resident set size (RSS) of each process is obtained from \texttt{ru\_maxrss}, and the table reports the maximum of the three measurements in MiB.
\end{tablenotes}
\end{threeparttable}
\end{table}

To distinguish the effects of the number of true-twin classes and the size of each class on compilation overhead, we also fix the quotient graph at $K_{2,6}$ with eight modules, increase the number of data qubits from 16 to 256, and measure single-candidate compilation overhead separately for the full-cache and source-level optimal cache schemes. Table I-7 in the Online Appendix gives the results. The sums of median stage times for the two paths increase from 0.26 and 0.28 seconds to 4.82 and 6.99 seconds, respectively. The median sampled peak RSS values for the 256-vertex instance are 162.9 MiB and 164.7 MiB, respectively.

\subsection{Numerical Correctness Checks}
To check the implementation correctness of specific gate sequences and backend transformations, we compare complex amplitudes, including global phase, on fixed small programs. The maximum error for complete programs with multiple sweeps is $2.06\times10^{-13}$, and the maximum error on selected feasible inputs after physical routing is $2.52\times10^{-14}$. The checks also cover local query kernels, translation into the native gate set, and wire-order restoration. Section C.6 of the Online Appendix gives the test conditions and error definitions.

\section{Conclusion}
\label{sec:con}

We address the repeated computation of overlapping neighborhood conditions in constraint-preserving mixer programs for graph independent sets by presenting a structure-aware compilation method based on persistent occupancy encoding. We exploit mutual exclusivity of occupancy within true-twin classes on the independent-set feasible domain to construct a reversible encoding on the original data register. This allows class occupancy information to be updated in step with legal local flips and maintained across ordered mixer operations and intervening cost layers. Using target compensation, blockwise parity queries, coherent restoration, and transformation of nonuniform cost coefficients, we prove that the complete compiled circuit has the same unitary action as the canonical program on the feasible subspace. Under a fixed circuit recipe, source-level caching cost is exactly separable with respect to the choice of true-twin classes. The amortization criterion for multiple mixer sweeps characterizes the relationship between reuse savings and encoding and decoding boundary costs, providing a computable source-level basis for cache selection.

Relative to the lowest CX count per program across six fixed reference implementations, our method reduces the total CX counts in the main evaluation and independent confirmation sets by 22.61\% and 19.87\%, respectively, with CX savings for every program in both sets. Relative to circuits with the same query synthesis mechanism and no persistent caching, persistent encoding further reduces the total CX counts by 4.10\% and 9.13\%, respectively. The controlled experiments on graph structure show that the allocation of true-twin class sizes affects the benefits of occupancy reuse across operations. Under the tested conditions, concentrating vertices in larger true-twin classes yields more pronounced incremental caching savings. In the separate physical routing experiment, including terminal restoration of the original wire order, total CX counts decrease by 9.14\% on the linear topology and by 12.50\% on the two-dimensional grid. Overall, turning the mutually exclusive occupancy of true-twin classes into a circuit representation that is continually updated across operations reduces the CX cost of complete circuits while preserving the coherent semantics of ordered programs.

\normalfont\normalsize
\doublespacing

\section*{Data availability}

The 12 public graph instances were obtained from the CMU COLOR benchmark collection \citep{cmucolor}, available at \url{https://mat.tepper.cmu.edu/COLOR/instances.html}.

The compiler source code, benchmark data, experimental configurations, circuit resource measurements, and scripts for reproducing the results are publicly available at \url{https://github.com/Moonooo24/persistent-occupancy-compiler}.

\section*{Acknowledgements}

Xiaoxing Liu acknowledges financial support from the Major Program of the National Social Science Fund of China (Grant No. 24\&ZD117). We thank the CMU COLOR maintainers and instance contributors for the benchmark graphs (\url{https://mat.tepper.cmu.edu/COLOR/instances.html}), the Qiskit contributors for circuit synthesis and routing tools (\url{https://github.com/Qiskit/qiskit}), the QuCLEAR authors for their baseline implementation (\url{https://github.com/revilooliver/QuCLEAR}), and the Cirq developers for the stabilizer code used in global-phase checks (\url{https://github.com/quantumlib/Cirq}).

\section*{Author contributions}

Zhiyi Wang: Conceptualization, methodology, software, formal analysis, investigation, and writing--original draft. Xiaorui Xue: Methodology, validation, visualization, and writing--review and editing. Xiaoxing Liu: Conceptualization, supervision, project administration, and writing--review and editing. All authors reviewed and approved the final manuscript.

\section*{Competing interests}

The authors declare no competing interests.

\section*{Declaration of generative AI and AI-assisted technologies in the manuscript preparation process}

During the preparation of this manuscript, the authors used ChatGPT (OpenAI) for language polishing and assistance in identifying potential algorithmic errors. The authors reviewed and verified the outputs and take full responsibility for the content of this work.

\clearpage
\singlespacing 
\setcounter{section}{0}
\setcounter{subsection}{0}
\setcounter{equation}{0}
\setcounter{table}{0}
\renewcommand{\thesection}{\Alph{section}}
\renewcommand{\theHsection}{appendix.\Alph{section}}
\numberwithin{equation}{section}
\renewcommand{\thetable}{I-\arabic{table}}
\renewcommand{\theHtable}{appendix.table.\arabic{table}}

\begin{center}
    \vspace*{1cm}
    {\Large\bfseries Online Appendix\par}
\end{center}

\noindent\begin{center}
\begin{minipage}{0.95\textwidth}
    \setlength{\parskip}{0.5em}
    \hyperref[app:proofs]{Appendix A. Proofs of Structural Encoding and Coherent Composition} \par
    \hyperref[app:kernels]{Appendix B. Gate-Level Constructions and Proofs for Coherent Query Kernels} \par
    \hyperref[app:repro]{Appendix C. Experiments, Implementation Protocols, and Verification} \par
    \hyperref[app:instances]{Appendix D. Supplementary Experimental Results and Per-Program Resources} \par
    \vspace{0.5em}
    \hyperref[tab:kernel-prefix]{Table I-1. Control-Computation Gate Sequences for Fixed-Rank Query Kernels} \par
    \hyperref[tab:high-rank]{Table I-2. Resources of High-Rank Phase Query Kernels and Full-Space Exact Paired Kernels} \par
    \hyperref[tab:reference-recipes]{Table I-3. Query Kernel Recipes for Reference Implementations, Ablations, and Our Method} \par
    \hyperref[tab:instance-confirm]{Table I-4. Per-Program Resources for the Independent Confirmation Set} \par
    \hyperref[tab:instance-public]{Table I-5. Resources for Public Graphs under Two Target Orders} \par
    \hyperref[tab:routing-ledger]{Table I-6. CX Resources for Physical Routing and Wire-Order Restoration} \par
    \hyperref[tab:single-path]{Table I-7. Time and Memory Overhead of Two Single-Candidate Paths} \par
\end{minipage}
\end{center}

\Needspace{12\baselineskip}
\refstepcounter{section}
\phantomsection
\section*{Appendix A. Proofs of Structural Encoding and Coherent Composition}
\label{app:proofs}
\addcontentsline{toc}{section}{Appendix A. Proofs of Structural Encoding and Coherent Composition}
This appendix provides complete proofs of the structural properties underlying persistent occupancy encoding, reversible target compensation, and coherent restoration. It also derives the cost-layer coefficient transformation, source-level cache separability, and the multiple-sweep amortization criterion used in Sections 3 and 4 of the main text.

\subsection{Closed-Neighborhood Equivalence Classes and Blockwise Control Rank}
\label{app:closed-neighborhood}
Equality of closed neighborhoods defines an equivalence relation on the vertex set. For distinct vertices $u,v\in C$ in the same class, $u\in N[u]=N[v]$, so $uv\in E_G$ and each class is a clique. Suppose an edge $uv$ connects distinct classes $C,D$, with $u\in C$ and $v\in D$. For every $u'\in C$, we have $v\in N[u']=N[u]$. Using $N[v']=N[v]$ for every $v'\in D$ then gives $u'v'\in E_G$ for all $u'\in C$ and $v'\in D$.
Thus, the edges between distinct classes form either a complete bipartite edge set or the empty set.

For target $i\in C$, the original neighborhood is the disjoint union of $O=C\setminus\{i\}$ and all external neighboring classes. When $O$ is nonempty, $O$ is a clique and is fully adjacent to each external neighboring class, so $O$ can be merged with a fixed external neighboring class into one clique block.
If $|C|=1$, then $r_i=\delta_C$. If $|C|>1$ and $\delta_C>0$, there are still $\delta_C$ blocks after merging. If $\delta_C=0$ and $|C|>1$, the remaining vertices of the target class form one block. These three cases establish Eq.~\eqref{eq:rank} of the main text for the control rank of target $i$.

Closed-neighborhood equivalence ensures these partitioning and occupancy properties. As a comparison, the two endpoints of a three-vertex path have the same open neighborhood but can both be selected, so their parity does not accurately indicate whether that vertex set is occupied. A general maximal-clique cover can also contain overlapping cliques, and adjacency between its cliques need not be complete. The closed-neighborhood equivalence partition used here gives mutual exclusion within each class and complete adjacency or nonadjacency between distinct classes.

\subsection{Reversible Occupancy Encoding and Target Row and Column Isolation}
\label{app:encoding-proof}
Let $x$ be the original data column vector over $\GF$. The encoding rows for a cached class $C$ are
\begin{equation}
 (E_\A x)_{a_C}=\bigoplus_{j\in C}x_j,\qquad (E_\A x)_j=x_j\quad(j\ne a_C).
\end{equation}
If target $i$ is a nonanchor vertex in a cached class, $K_i$ removes $x_i$ from the anchor row. The target itself has a unit row, and no other row contains $x_i$, so column $i$ is also a unit column. If the target is the anchor, $K_i$ fully undoes the class's star-shaped CX encoding. If the target class is uncached, both its encoding and compensation are identities. The supports of the other classes are disjoint from the target class and leave these conclusions unchanged.

The temporary data coordinates satisfy $y=T_ix$, and the query condition satisfies $fy=bx$. The query row is therefore determined by the inverse of the reversible transformation, namely $f=bT_i^{-1}$.

The support of each original neighborhood block in temporary coordinates can be decomposed into the following class components. The support of an entire uncached external class $D$ remains $D$, whereas the support of an entire cached external class is $\{a_D\}$. For an uncached target class, the set $O$ of remaining vertices is unchanged. For a cached class with a nonanchor target, its support is $\{a_C\}$, and for a cached class with an anchor target, its support remains $O$. Merging takes only the union of two previously disjoint class components. Thus, the nonzero supports of all parity query rows are nonempty, pairwise disjoint, and exclude the target bit.
The supports are disjoint and nonempty, so the rank of the query matrix equals the number of neighborhood blocks.

\subsection{Coherent Restoration after a Legal Flip and Program Composition}
\label{app:recovery-proof}
Let $S_i=M_iK_iE_\A$. Its target row and column are still the unit vector $e_i$. On an original feasible state, the occupancy of each true-twin class is at most 1, so each pivot bit equals the occupancy flag of the corresponding neighborhood block. When $x_i=1$, all neighborhood blocks are empty, and the input is in the admissible input domain defined by Eq.~\eqref{eq:care} of the main text. All pivots are zero if and only if the canonical neighborhood NOR condition holds.

When the neighborhood zero-control condition is not satisfied, the query kernel acts as the identity on the corresponding admissible branch, and the complete macro outputs $E_\A x$. When the condition is satisfied, the action of the query kernel in temporary coordinates is
\begin{equation}
 \cos\beta\ket{S_ix}-\ii\sin\beta\ket{S_ix\oplus e_i}.
\end{equation}
From $S_i^{-1}e_i=e_i$ and $S_i^{-1}=E_\A^{-1}K_i^{-1}M_i^{-1}$, we have
\begin{equation}
 K_i^{-1}M_i^{-1}e_i=E_\A e_i.
\end{equation}
After restoring the data coordinates with the inverse of the actual preparation transformation, the resulting state is
\begin{equation}
 \cos\beta\ket{E_\A x}-\ii\sin\beta\ket{E_\A(x\oplus e_i)}.
\end{equation}
Both original states satisfy the independent-set constraints, and the coefficients agree with the canonical $R_x(2\beta)$ rotation. Linear extension from the feasible computational basis states gives the operator identity on the entire feasible subspace, which continues to hold after tensoring with any reference system. For a nonanchor target in a cached class, $E_\A e_i$ includes updates to both the target and anchor bits, and the local inverse transformation keeps the target flip consistent with the occupancy representation.

\subsection{Nonuniform Costs and Global Phase Consistency}
\label{app:cost-proof}
On a cached class $C$, at most one original variable in the same true-twin class is 1, so the parity sum over the binary field equals the corresponding integer sum. Hence
$y_a=\sum_{j\in C}x_j$ and $y_j=x_j$ for $j\ne a$. It follows that $x_a=y_a-\sum_{j\in C\setminus\{a\}}y_j$. Substituting into the original cost expression $\sum_{j\in C}w_jx_j$ yields Eq.~\eqref{eq:cost} of the main text. For uncached classes, the cost layer temporarily computes clique occupancy parity and restores the original data coordinates through the actual inverse after applying the phase operation.

If the terms of the actual representation are $c_s\ell_s(y)$, where $\ell_s$ is a single-bit variable or parity function, then the global phase factor for each term is $e^{\ii\gamma c_s/2}$ and the total factor is $\exp(\ii\gamma\sum_s c_s/2)$. The coefficient sums for different representations can differ from the sum of the original weights.
For example, for a three-vertex clique with weights $(2,-3,1)$, the original weight sum is 0. If the vertex with weight 2 is chosen as the anchor, the representation coefficients are $(2,-5,-1)$ and their sum is $-4$. The corresponding global phase factor $e^{-2\ii\gamma}$ must therefore be included in the complete circuit.

The compiler records the symbolic coefficients of the global phase in sparse affine form, extracts the constant coefficient for each independent parameter, and reconstructs gate angles and the global phase from the original expressions. Conversion from exact rationals to floating-point numbers introduces numerical representation error. Independent complex-amplitude comparisons therefore check the numerical agreement of the complete circuits at specified parameters.

\subsection{Cache Subset Separability and Amortization over Multiple Mixer Sweeps}
\label{app:amortization-proof}
Under a fixed source-level circuit recipe, the lowest-index physical support position for an entire external class component is always its anchor. Compensation for the target module also leaves the lowest-index support position of that block fixed. The pivot of a merged block is the smaller of the two fixed minima, so control wire ordering and the physical mapping of the query kernel are independent of the cache states of the other modules.

The endpoint multiplicities of each variable CX gate can be attributed to a unique cache class. Boundary encoding and target compensation depend only on their own class, and parity-preparation instructions on nonanchor control wires vary independently by class. Controls in open Gray synthesis vary by the same rule. Data wires belonging to different classes are disjoint, so the vector of source-level CX endpoint multiplicities is an affine function of the class cache indicators. Applying arbitrary prespecified nonnegative gate weights to this vector gives Eq.~\eqref{eq:affine} of the main text.

Encoding and decoding at the program boundaries each require $B(\A)$ source-level CX gates. The source-level costs of the $p$ mixer sweeps and $p-1$ intervening cost layers are $pF(\A)$ and $(p-1)H(\A)$, respectively, yielding Eq.~\eqref{eq:amortization} of the main text. If $\Delta_F+\Delta_H>0$, the smallest positive integer sweep count satisfying the strict savings criterion is
\begin{equation}
 p_{\min}=\max\left\{1,\left\lfloor\frac{2B(\A)+\Delta_H}{\Delta_F+\Delta_H}\right\rfloor+1\right\}.
\end{equation}
When $\Delta_F+\Delta_H\le0$, use Eq.~\eqref{eq:breakeven} of the main text directly to determine the net savings for each sweep count. In the absence of intervening cost layers, the criterion reduces to $p\Delta_F>2B$. The derivation applies to a fixed source-level circuit recipe. Changes in cost due to backend gate cancellation and physical routing are evaluated on the complete candidate circuits.

\Needspace{12\baselineskip}
\refstepcounter{section}
\phantomsection
\section*{Appendix B. Gate-Level Constructions and Proofs for Coherent Query Kernels}
\label{app:kernels}
\addcontentsline{toc}{section}{Appendix B. Gate-Level Constructions and Proofs for Coherent Query Kernels}
This appendix provides gate-level constructions and correctness proofs for the conditional query kernels described in Section 4.4 of the main text. Table~\ref{tab:kernel-prefix} lists the fixed-rank control-computation sequences (ranks 4--8), and Table~\ref{tab:high-rank} reports resource bounds and compiled resources for higher control ranks.
This appendix gives the gate-level definitions, residual-phase proofs, and analytical resource bounds for the open Gray--Walsh circuit and conditional query kernel. The main target is indexed by 0, and the ordered controls by $1,\ldots,r$. Gate sequences are written and executed in temporal order from left to right.
Define $R_x(\theta)=e^{-\ii\theta X/2}$, $R_z(\theta)=e^{-\ii\theta Z/2}$, and $T=\operatorname{diag}(1,e^{\ii\pi/4})$.

\subsection{Open Gray--Walsh Circuit and Terminal Frame Phase}
\label{app:gray-frame}
For disjoint query rows $f_1,\ldots,f_r$, all XOR combinations are distinct, so Eq.~\eqref{eq:walsh} of the main text contains $2^r$ terms. Let $g(k)=k\oplus\lfloor k/2\rfloor$, $k=0,\ldots,2^r-1$, and compute $m_k=\bigoplus_{b:g(k)_b=1}f_b$ in a fixed factor order.
Starting with $m_0=0$, for each $k\ge1$, apply CX gates in ascending qubit order at the nonzero positions of $m_{k-1}\oplus m_k$, using the corresponding control bits as controls and the target bit as the target. This updates the accumulated parity. Each rotation term is implemented through a target Hadamard transformation and $R_z(2\beta/2^r)$.
The closed version undoes the control transformation for the current binary mask $f_*$ at the end, while the open version retains this terminal frame. The phase difference operator between the two after target Hadamard conjugation is
\begin{equation}
 R_* =\prod_{j\in\supp f_*}\mathrm{CZ}_{ij}.
\end{equation}
$f_*T_i$ corresponds to the union of several original neighborhood blocks, and its nonzero positions lie in $N(i)$. When the target is zero, all CZ gates above act as the identity. When the target is one, independent-set feasibility forces every relevant neighborhood bit to zero. Thus, the residual operator is the identity on the feasible output space of the complete mixer macro. This property is verified at the macro output boundary, and the internal temporary data coordinates are determined by the corresponding reversible transformations.

\subsection{Three-CX Relative-Phase Toffoli Primitive}
\label{app:toffoli-word}
Let $\tau(a,b;c)$ denote the following gate sequence:
\begin{equation}
\label{eq:tau-word}
[H_c,T_c,\CX(b,c),T_c^\dagger,\CX(a,c),T_c,
 \CX(b,c),T_c^\dagger,H_c].
\end{equation}
For fixed $(a,b)$, the gate applies $I$ on the target wire when $a=0$, $Z$ when $(a,b)=(1,0)$, and $Y$ when $(a,b)=(1,1)$. Its computational-basis permutation therefore agrees with CCX, although the basis-state phases may differ, and it uses 3 CX gates.
The control-computation prefixes below act only on the control register and are followed by their actual inverses. The intervening flip of the main target uses an exact CCX, whose fixed decomposition contains 6 CX gates. The phases cancel exactly on every computational-basis state of the controls through the computation prefix and its actual inverse.

\subsection{Two-Control Phase Query Subkernel \texorpdfstring{$B_2$}{B2}}\label{app:two-control}
Given controls $u,v$ and target $t$, define
\begin{equation}
\label{eq:B2-word}
\begin{split}
B_2(\delta)=\operatorname{seq}[H_t,R_{z,t}(\delta/2),\CX(u,t),R_{z,t}(\delta/2),\\
\CX(v,t),R_{z,t}(\delta/2),\CX(u,t),R_{z,t}(\delta/2),H_t].
\end{split}
\end{equation}
The identity on the full local space is
\begin{equation}
\label{eq:B2-completed}
 B_2(\delta)=U_{\NOR,2}(\delta)\mathrm{CZ}_{tv}
 =R_{x,t}(2\delta\NOR_{\{u,v\}})Z_t^v.
\end{equation}
Adding the control transformation $\CX(v,t)$ at the end closes the phase frame. The four equally weighted parity rotations then combine into the Walsh expansion of the zero-control projector, corresponding to the complete NOR-controlled rotation. After target Hadamard conjugation, the terminal frame is $\mathrm{CZ}_{tv}$. This operator is the identity for $u=v=0$ and therefore commutes with the NOR-controlled rotation.
The gate sequence contains 3 CX gates and 4 parameterized rotations.

The three-CX subkernel used here can also be derived by an equivalent two-step factorization. First, let the single-control kernel be
\begin{equation}
B_1(\delta)=\operatorname{seq}[H_t,R_z(\delta),\CX(u,t),R_z(\delta),H_t].
\end{equation}
Partitioning by $u=0,1$ gives $B_1=U_{\NOR,1}(\delta)Z_t^u$. Using $X_t^{1-v}$ as the exact flip, form the following product
$B_1(\delta/2)Z_t^{1-v}B_1(-\delta/2)Z_t$. The two residual involutions $Z_t^u$ cancel, and the resulting complete unitary action agrees with the two-control subkernel above.

\subsection{Gate-Level Constructions of Rank-4 through Rank-8 Query Kernels}
\label{app:finite-kernels}
Use $L=\{1,2\}$ and $R=\{3,\ldots,r\}$ throughout. Let $F_R=\prod_{j\in R}X_j$ be the open-control flips. Table~\ref{tab:kernel-prefix} specifies the computation prefix $V_r$ and the two controls $a_r,b_r$ of the exact CCX. Define
\begin{equation}
\label{eq:Ar-specific}
 A_r=F_R V_r^\dagger\mathrm{CCX}(a_r,b_r;0)V_r F_R,
\end{equation}
With matrices acting from right to left, the complete query kernel is
\begin{equation}
\label{eq:finite-production-kernel}
 \kappa_r(\beta)=B_2(\beta/2)H_0 A_r H_0 B_2(-\beta/2)Z_0.
\end{equation}
The two-control subkernel can be written as $U_{\NOR,2}(\delta)D_L$, where $D_L=Z_0^{f(L)}$, $f(L)\in\{0,1\}$, $f(0)=0$, and $D_L$ is parameter independent. This involution is compatible with the conditional query and cancels in the composite kernel by the algebraic relation above. The resource counts in the evaluation correspond to gate-level implementations fixed before the experiments.
The control-computation prefix acts only on the control register, and $B_2$ always acts on the main target and original controls 1 and 2.

Let $q_r(z)$ be the Boolean control condition of the complete flip $A_r$ with no residual phase. For $r=4,5,6$, the corresponding computation prefixes form AND-computation chains with 0, 1, and 2 conditionally available control wires as workspace, respectively. They satisfy $q_r=\NOR_R$ when $L=0$.
When $L\ne0$, the values of $q_r$ in the remaining control sectors are determined by the computation prefix of the specific completion unitary. The actual inverse of this prefix restores all control bits.

For $r=7$, denote the bits of $R$ after open-control conversion by $a,b,c,d,e$. On $L=0$, the Boolean values of the two intermediate controls after the computation prefix are $ab$ and
$1\oplus a\oplus e(1\oplus b\oplus cd)$, respectively. If $ab=0$, then $q_7=0$. If $ab=1$, the latter is $ecd$, so $q_7=abcde$. For $r=8$, write the bits after open-control conversion as $a,b,c,d,e,f$. The second intermediate control is
$1\oplus a\oplus(1\oplus d\oplus ef)(1\oplus b\oplus cd)$.
When $ab=1$, it reduces to $cd(1\oplus d\oplus ef)=cdef$, so $q_8=abcdef$. This proves that all listed prefixes satisfy $L=0\Rightarrow q_r=\NOR_R$.

Substituting into Eqs.~\eqref{eq:B2-completed} and \eqref{eq:finite-production-kernel} yields
$\kappa_r(\beta)=U_{\NOR,r}(\beta)Z_0^{1-q_r}$.
The residual involution acts as the identity on branches with a zero target and on branches with all controls zero. It satisfies the query and restoration contract defined in the main text and guarantees that pairing two half-angle completion kernels is exact on the full space. The tabulated CX counts sum the relative-phase computation prefix and its actual inverse, one exact CCX, and two three-CX subkernels. The analytical count is $6r-12$.

For control rank $r>8$, retain the factorization with $|L|=2$ and use the KG conditionally clean synthesis~\citep{khattar} to generate the control-computation prefix. This construction satisfies the coherent restoration condition, and its CX resource bound grows linearly in the control rank.

\begin{table}[H]
\centering
\caption{Control-computation gate sequences for fixed-rank query kernels}
\label{tab:kernel-prefix}

\begin{threeparttable}
\scriptsize
\setlength{\tabcolsep}{4pt}
\renewcommand{\arraystretch}{1.05}

\begin{tabular*}{\textwidth}{@{\extracolsep{\fill}}c p{.58\linewidth} cc@{}}
\toprule
$r$ & Gate sequence of $V_r$ & $(a_r,b_r)$ & CX count\\
\midrule
4 & $\varnothing$ & $(3,4)$ &12\\
5 & $\tau_{34;1}$ & $(1,5)$ &18\\
6 & $\tau_{34;1};\tau_{15;2}$ & $(2,6)$ &24\\
7 & $\tau_{34;1};\tau_{56;4};X_4;\tau_{74;3};X_3$ & $(1,3)$ &30\\
8 & $\tau_{34;1};\tau_{56;4};X_4;\tau_{78;6};X_6;$\newline
$\tau_{64;3};X_3$ & $(1,3)$ &36\\
\bottomrule
\end{tabular*}

\begin{tablenotes}[flushleft]
\scriptsize
\item $\tau_{ab;c}$ denotes the relative-phase Toffoli gate defined in Eq.~\eqref{eq:tau-word}. Semicolons concatenate gates in execution order, and $\varnothing$ denotes the empty sequence. The constructed query kernels act only on the original target and the $r$ control wires.
\end{tablenotes}
\end{threeparttable}
\end{table}

\subsection{Deterministic Implementation of High-Rank Conditional Query Kernels}
\label{app:high-rank}
For $r\ge9$, the computation prefix uses the conditionally clean multi-controlled synthesis primitive \nolinkurl{synth_mcx_1_clean_kg24} in Qiskit 2.1.2. Here $m=r-2$, and the primitive's input wire order comprises $m$ positive control wires, one target wire, and one workspace wire satisfying the cleanliness precondition in the specified sector.
For ordered controls $1,\ldots,r$, the following deterministic wire mapping and gate-sequence extraction yield $V_r$ in Eq.~\eqref{eq:Ar-specific}, from which the phase kernel is constructed.
\begin{enumerate}
\item Take $L=(1,2)$ and $R=(3,\ldots,r)$. Map the SDK circuit's wire indices in order to $(R,0,1)$, where 0 is the main target and original control 1 is the workspace. No separate ancillary data qubit is added.
\item Locate the unique exact \texttt{ccx} in the gate sequence. Define $V_r$ as all gates preceding it in execution order. Verify that each gate has type \texttt{x}, \texttt{rccx}, or \texttt{rccx\_dg} and that none acts on the main target after mapping. Read the two mapped controls of the exact \texttt{ccx} and verify that its target is 0.
\item Execute $F_R,V_r,\mathrm{CCX},V_r^\dagger,F_R$. Obtain $V_r^\dagger$ by taking the actual inverse of the prefix circuit, executing the inverse gates in reverse order. Combine this flip with two invocations of $B_2$ according to Eq.~\eqref{eq:finite-production-kernel}. For a full-space exact kernel, apply two completion kernels in sequence, each with half the target angle, as in Eq.~\eqref{eq:half-pair} of the main text.
\end{enumerate}
On the $L=0$ subspace, the workspace mapped to original control 1 satisfies the conditional cleanliness precondition of the KG primitive, and the corresponding flip condition is $\NOR_R$. In the other control sectors, the computation prefix is a phased permutation on the control computational basis. Combining an exact CCX with the actual inverse of the prefix gives a conditional permutation without control-basis-dependent residual phases and restores all control bits.
The two two-control subkernels and target compensation determine the residual phase of the completion unitary. Its admissible input domain is specified by Eq.~\eqref{eq:care} of the main text. The prefix contains at most $m-2$ three-CX relative-phase Toffoli gates. Including its actual inverse, the six-CX exact CCX, and the two three-CX subkernels gives the construction upper bound $6(m-2)+6+6=6r-12$.

\subsection{Residual Phase Involution, Exact Half-Angle Pairing, and Resource Bounds}
\label{app:kernel-proof}
For the conditional phase query kernel with $r\ge4$, this section derives the full-space residual phase and an analytical CX resource bound.
Partition the controls $N$ into a fixed set $L$ of two control wires and $R=N\setminus L$, and let $s(z)=\NOR_L(z)$.
For $r=4$, $|R|=2$, and an exact CCX is applied directly after converting the controls on $R$ to open controls.
For $5\le r\le8$, use the fixed computation prefixes specified in Table~\ref{tab:kernel-prefix}.
For $r\ge9$, use the KG one-clean-ancilla synthesis recipe (one-clean) to construct the computation prefix and map its conditionally clean workspace wire to original control wire~1 in $L$.
The cleanliness precondition of the conditionally clean primitive is imposed only on the $L=0$ subspace.

Let $\mathcal C$ include the open-control $X$ flips on $R$ and this computation prefix.
$\mathcal C$ acts only on control data and is a phased permutation on the computational basis.
Apply an exact CCX in the middle, followed by the actual inverse of $\mathcal C$.
Thus, the computation phases in $\mathcal C^\dagger\mathrm{CCX}\mathcal C$
cancel exactly for every initial control value, giving
$A\ket{t,z}=\ket{t\oplus q(z),z}$.
Here $q$ is a Boolean function independent of the main target bit and the rotation parameter.
The actual inverse coherently restores all control data, and the above construction guarantees the required condition
\begin{equation}
 s(z)=1\ \Longrightarrow\ q(z)=\NOR_R(z),
 \qquad q(0^r)=1.
\end{equation}
When $L\ne0$, $q$ is determined by the control computation of the actual completion unitary.

The fixed two-control subkernel must satisfy the completion-unitary condition on the entire local target--control space
\begin{equation}
 B(\delta)=R_{x,i}(2\delta s)D_B,\qquad
 D_B=Z_i^{f(L)},\quad f(L)\in\{0,1\},\quad f(0^2)=0.
\end{equation}

The same subkernel has the same parameter-independent $D_B$ for both positive and negative parameter values.
It commutes with $Z_i^{q(z)}$ and with $R_{x,i}(\alpha s)$,
since the latter is the identity for $s=0$, while $D_B=I$ for $s=1$.

With matrices acting from right to left, the original construction is
\begin{align}
 \kappa_r(\beta)
 &=B(\beta/2)Z_i^{q(z)}B(-\beta/2)Z_i\\
 &=R_{x,i}(\beta s)Z_i^{q(z)}
   R_{x,i}(-\beta s)Z_i\\
 &=R_{x,i}(2\beta s q)Z_i^{1-q(z)}.
\end{align}
Since $s q=\NOR_L\NOR_R=\NOR_N$,
setting $D_r=Z_i^{1-q(z)}$ gives Eq.~\eqref{eq:kernel-involution} of the main text.
$D_r$ is a parameter-independent diagonal involution that acts as the identity on the NOR-enabled sector,
so it commutes with $U_{\NOR,r}(\beta)$. It also acts as the identity on zero-target branches.
The involution and commutation properties of the residual operator therefore give the following identity for every real $\beta$ on the entire target and control registers:
\begin{equation}
 \kappa_r(\beta/2)^2
 =U_{\NOR,r}(\beta/2)^2D_r^2
 =U_{\NOR,r}(\beta)
\end{equation}

The construction uses only the target and original control qubits, with original control wires providing the required conditional workspace.
Let $k=|R|=r-2\ge2$.
The computation prefix and its actual inverse use at most $2(k-2)$ three-CX relative-phase Toffoli gates in total.
The exact CCX uses 6 CX gates, and the two fixed two-control subkernels each use 3 CX gates.
For $k=2$, the computation prefix is empty. The analytical CX upper bound is therefore
\begin{equation}
 3\cdot2(k-2)+6+2\cdot3=6r-12,
 \qquad \mathrm{CX}_{\mathrm{exact\ pair}}\le12r-24.
\end{equation}
This analytical bound applies to the conditional phase construction for $r\ge4$. The experiments use this scalable implementation for $r>8$. Lower-rank primitives use their respective prespecified recipes and satisfy the same coherent restoration contract. The evaluation reports analytical bounds, measured compilation counts, and topology routing costs separately.

Table~\ref{tab:high-rank} compares the resource costs of the same construction under the phase-kernel and full-space exact-kernel specifications. Ranks 9 and 10 also undergo independent complex-amplitude verification. Ranks 12 and 16 report the measured compilation resources of the corresponding gate-level constructions.

\begin{table}[H]
\centering
\caption{Resources of high-rank phase query kernels and full-space exact paired kernels}
\label{tab:high-rank}

\begin{threeparttable}
\scriptsize
\setlength{\tabcolsep}{4pt}
\renewcommand{\arraystretch}{1.05}

\begin{tabular*}{\textwidth}{@{\extracolsep{\fill}}crrrr@{}}
\toprule
$r$ & Analytical CX bound & Native CX & Post-processed CX & Param. rotations\\
\midrule
9  & 42/84  & 42/84  & 42/84  & 8/16\\
10 & 48/96  & 48/96  & 48/96  & 8/16\\
12 & 60/120 & 60/120 & 60/120 & 8/16\\
16 & 84/168 & 84/168 & 84/168 & 8/16\\
\bottomrule
\end{tabular*}

\begin{tablenotes}[flushleft]
\scriptsize
\item Each pair $a/b$ denotes a single phase kernel and an exact kernel formed by two half-angle phase kernels in sequence, respectively. The table reports analytical CX upper bounds, native CX counts, post-processed CX counts, and parameterized rotation counts.
\end{tablenotes}
\end{threeparttable}
\end{table}

\Needspace{12\baselineskip}
\refstepcounter{section}
\phantomsection
\section*{Appendix C. Experiments, Implementation Protocols, and Verification}
\label{app:repro}
\addcontentsline{toc}{section}{Appendix C. Experiments, Implementation Protocols, and Verification}
This appendix documents the graph and program generation rules, random seeds, fixed reference implementations, topology routing, software environment, and numerical verification used in Section 5 of the main text. Table~\ref{tab:reference-recipes} summarizes the query-kernel recipes for the reference circuits, ablations, and the proposed method.
This appendix describes the graph and program generation protocols, fixed reference circuit recipes, implementation environment, and correctness verification procedure.

\subsection{Data Groups, Instance Generation, and Random Seeds}
\label{app:data-protocol}
Instance generation for the main evaluation set covers different data sizes, regular quotient-graph degrees, and true-twin class size patterns. Under the specified generation configurations, these factors jointly determine the range of query control ranks. Deduplicating graphs by labeled vertices and undirected edge sets yields 62 graphs. Deduplicating by graph, number of mixer sweeps, weight configuration, and actual target permutation then yields 134 programs from 144 original program records. The independent confirmation set uses separate size configurations and random generation streams. A hash-based prefilter and exact isomorphism checks on quotient graphs that retain module-size attributes yield 36 nonisomorphic graphs and 72 programs.
All 36 graphs satisfy the connectivity and isomorphism screening conditions on the first generation attempt. Sample acceptance for the independent confirmation set is based solely on connectivity and quotient-graph isomorphism checks that retain module-size attributes.

The public graphs are recorded with their original data, source URLs, file headers, and normalized undirected edge sets. Edges repeated in opposite directions are treated as the same undirected constraint. Fixed small graphs are used for numerical consistency checks of specific circuits, and scaling graphs are used to measure stage times and memory.

Input graphs in both the main and confirmation sets are constructed by replacing regular quotient-graph vertices with true-twin cliques. Vertices are assigned consecutive identifiers within each class. Each class is a clique, and adjacent classes are fully connected. The initial seed for the main evaluation set is $202611090000+10000n+100d+30I_{\rm mixed}+j$, where $d$ is the quotient-graph degree and $j=0,1,2$ indexes the generation repetitions. The generator tries successive incremented seeds until it obtains a graph that meets the connectivity requirement.
The candidate seed for the confirmation set is $2026111000000+1000003i+15485863a$, where $i$ indexes the fixed generation order and $a$ is the retry index. Acceptance is based on connectivity and exact isomorphism screening that preserves module sizes. For both sets, the seed for the fixed pseudorandom target order is the corresponding graph generation seed plus 701. The canonical input records store the actual graph, order, and retry index.

The graphs and ordered programs in the controlled structure experiments are fixed before resource counting. The five coverage levels in the first series use (number of singleton classes, number of two-vertex classes) equal to $(32,0)$, $(24,4)$, $(16,8)$, $(8,12)$, and $(0,16)$, respectively. There are two fixed placements for each intermediate level and one for each endpoint, for a total of 8 graphs.
The second series fixes 16 quotient-graph vertices and uses four class-size distributions: 15 singleton classes and one 17-vertex class, 12 singleton classes and four five-vertex classes, 8 singleton classes and eight three-vertex classes, and 4 singleton classes, eight two-vertex classes, and four three-vertex classes. Their coverages are 53.125\%, 62.5\%, 75\%, and 87.5\%, respectively.
The expanded classes in each quotient graph are placed according to a fixed pseudorandom permutation, with placement seed $202611110+j$, where $j$ is the placement identifier. The seed for the fixed pseudorandom target order is 202611117. All input vertex weights are 1, all sweeps on a given graph use the same order, and the $p-1$ cost-layer parameters are independent. The two series contain 12 graphs and 72 programs in total. Identical cache subsets are deduplicated for resource measurement.

\subsection{Reference Implementations and Fixed Recipes}
\label{app:reference-recipes}
Table~\ref{tab:reference-recipes} lists implementations of prior constructions, the fixed full-space exact reference, and the high-rank query implementations shared by the comparison methods. The control rank is at most 8 in both the main evaluation and independent confirmation sets. Both complete comparisons use finite-rank recipes fixed before the experiments.

The specific recipe is selected by control rank. For $r=1,2$, closed Walsh synthesis uses 2 and 4 CX gates, respectively. For $r=3$, a phase-independent paired-query construction and a two-control partial subkernel use 8 CX gates.
For $r=4$, pairing a relative-phase intermediate Toffoli with an exact two-control subkernel uses 14 CX gates.
For $r=5,6$, paired RC3X constructions with $|L|=2,3$, respectively, use 20 and 28 CX gates.
For $r=7,8$, fixed Iten exact flips paired with partial subkernels with $|L|=4,5$ and $|R|=3$ use 40 and 52 CX gates, respectively.
These recipes are defined by prespecified blockwise pairings and gate-level primitives. Each exact macro implements the specified NOR-controlled rotation on the entire local space. The original cost and clique cost are measured as separate reference implementations to keep the comparison configurations explicit.

The matched exact pairing can also be constructed from the following operator identities. Let $E_L(\theta)$ denote the full-space exact rotation with zero-valued controls on $L$,
and let $\mathcal Z_L(\theta)=H_iE_L(\theta)H_i$. Let $A_R$ be a monomial flip with positive controls on $R$. It may have diagonal phases but must restore all other data. Its computational-basis permutation gives
$A_R^\dagger Z_i A_R=(-1)^{\prod_{j\in R}x_j}Z_i$. The diagonal phases cancel under this conjugation.
Use $F_R$ to convert the controls in $R$ to zero controls. The paired construction is
\begin{equation}
E_{L,R}(\beta)=F_R H_i\mathcal Z_L(\beta/2)A_R^\dagger
\mathcal Z_L(-\beta/2)A_RH_iF_R.
\end{equation}
Combining the two exponentials gives $e^{-\ii\beta Z_i\NOR_L\NOR_R}$. Hadamard conjugation on the target then gives the complete NOR-controlled rotation.
When $A_R$ is an exact flip without additional phases, a partial subkernel with $C_L=U_LD_L$, $D_L=Z_i^{f(L)}$, and $f(0)=0$ can also be used. The residual after conjugation is $X_i^{f(L)}$, which commutes with $A_R$ and cancels between the two subkernels.
For $r=3,7,8$, the table uses an exact flip with no additional phases, paired with a specified partial subkernel. The remaining high-rank exact kernels combine relative-phase flips with exact subkernels.

\begin{table}[H]
\centering
\caption{Query kernel recipes for reference implementations, ablation configurations, and our method}
\label{tab:reference-recipes}

\begin{threeparttable}
\scriptsize
\setlength{\tabcolsep}{4pt}
\renewcommand{\arraystretch}{1.05}

\begin{tabular*}{\textwidth}{@{\extracolsep{\fill}}p{.22\linewidth}p{.32\linewidth}p{.34\linewidth}@{}}
\toprule
Implementation or configuration & $r\le8$ & $r>8$\\
\midrule
Structured/Vale & Smallest-pivot CPG, fixed Vale rotation, and actual inverse & Same CPG with a fixed ancilla-free exact double-query implementation in Qiskit\\
Closed CPG & Full Walsh synthesis in target Gray order with a closed frame & Parity preparation, fixed ancilla-free exact double query, and actual inverse\\
Open CPG (including the full Walsh ablation) & Open Walsh synthesis in target Gray order & Shared high-rank conditional phase query kernel\\
Matched exact kernel & Fixed full-space exact query kernel library & Exact pairing of two half-angle phase kernels of the same construction\\
Our method & Open Walsh for $r\le3$ and Table~\ref{tab:kernel-prefix} kernels for $r\ge4$ & High-rank conditional phase query kernel\\
\bottomrule
\end{tabular*}

\begin{tablenotes}[flushleft]
\scriptsize
\item All implementations use the same number of data qubits and retain the terminal operations required for coherent correctness.
\end{tablenotes}
\end{threeparttable}
\end{table}

\subsection{Topology Routing and Restoration of the Original Wire Order}
\label{app:routing-method}
The linear topology consists of edges $(i,i+1)$ between adjacent vertices. For the two-dimensional grid, let $a$ be the largest positive divisor of the graph size no greater than $\sqrt n$, and construct an $a\times(n/a)$ rectangular coupling grid. Device coupling edges support CX gates in both directions. The initial layout is the identity mapping, and after routing the compiler backend provides the permutation from logical to physical qubits.
To restore the original physical wire order of the logical data, we select deterministic shortest paths on the coupling graph. Swaps between adjacent positions on the path and any necessary reverse swaps transpose the endpoints while returning the interior logical qubits to their positions before the endpoint transposition. The logical-to-physical mapping is updated after each swap, and the identity permutation is restored when all swaps are complete.

Each SWAP is decomposed into three CX gates that obey device connectivity. The final circuit undergoes a common bounded number of local simplification rounds, and both the physical edge constraints on two-qubit gates and the original logical wire order are checked. The three routing seeds are 20261109, 20261110, and 20261111.
The physical evaluation includes 22 ordered programs. Sixteen come from two-vertex-class and mixed-class instances in the main evaluation set with $n\in\{32,64\}$ and $d\in\{4,6\}$, under the two target orders. The other six come from two orders for each of three fixed mechanism graphs: a 14-vertex H7 graph, a 12-vertex graph obtained by replacing each vertex of the bipartite quotient graph $K_{2,4}$ with a two-vertex clique, and a 14-vertex graph obtained by the same replacement procedure from $K_{2,5}$.
The reproduction data record the complete inputs and graph identifiers. Numerical consistency is also checked after routing.

\subsection{Implementation Environment}
\label{app:environment}
The primary runtime environment uses Python 3.12.3, Qiskit 2.1.2, NumPy 1.26.4, SymPy 1.14.0, and mpmath 1.3.0. Timing is performed in independent processes and excludes runtime import overhead. The Linux memory metric in Table~\ref{tab:full-entry} of the main text is the peak resident set size (RSS) of the process,
read from \texttt{ru\_maxrss} and reported in MiB.

\subsection{Input Conditions and Structural Records}
\label{app:structure-cert}
An input program includes a simple undirected graph with vertices labeled $0,\ldots,n-1$ (no self-loops or parallel edges), $p\ge1$ mixer sweeps, a complete vertex permutation for each sweep, and decimal weights parsed as exact rationals. The compilation record stores the closed-neighborhood partition, quotient graph, temporary target transformation and its inverse, neighborhood blocks, and query rows so that the specific circuit can be reconstructed. The same construction rules apply to singleton classes, isolated vertices, and disconnected graphs.
The comparisons distinguish invalid inputs, unsupported implementations or primitives, and incomplete resource calculations. Resource comparisons include circuits that actually complete under matching program specifications.

\subsection{Numerical Correctness and Implementation Consistency Checks}
\label{app:verification}
We compare the full matrices of local kernels column by column, including global phase, covering the relative-phase Toffoli primitive, the two-control subkernel, and the fixed kernels of ranks 4 through 8. The rank-9 and rank-10 phase query kernels and their exact paired kernels are independently checked. For small complete programs, we compare the canonical ordered evolution with the compiled circuit using independent layer parameter values. The tested angles include zero and nonzero values of different signs and magnitudes. Numerical checks cover the original construction, native gate set translation, and bounded post-processing.

For the actual output state $\psi^{\mathrm{actual}}$ and the canonical reference state $\psi^{\mathrm{reference}}$, we use the complex-amplitude error including global phase
\begin{equation}
  \epsilon_\infty=\max_j\left|\psi_j^{\mathrm{actual}}-\psi_j^{\mathrm{reference}}\right|.
\end{equation}
The maximum error for the rank-9 and rank-10 local kernels is $4.10\times10^{-14}$, and for complete programs with multiple sweeps it is $2.06\times10^{-13}$. The same comparison on fixed feasible inputs after routing gives a maximum error of $2.52\times10^{-14}$. The full-space regression against the routing input circuits gives a maximum error of $9.67\times10^{-17}$.

\Needspace{12\baselineskip}
\refstepcounter{section}
\phantomsection
\section*{Appendix D. Supplementary Experimental Results and Per-Program Resources}
\label{app:instances}
\addcontentsline{toc}{section}{Appendix D. Supplementary Experimental Results and Per-Program Resources}
This appendix supplies detailed resource records and supplementary experimental comparisons for Section 5 of the main text. Table~\ref{tab:instance-confirm} lists all 72 independent-confirmation programs; Table~\ref{tab:instance-public} reports the two target orders on 12 public graphs; Table~\ref{tab:routing-ledger} separates physical routing and terminal restoration costs; and Table~\ref{tab:single-path} reports classical time and memory overhead.
This appendix reports the completion coverage of additional references, per-program resources for the independent confirmation set and public graphs, physical routing details, and single-candidate compilation overhead at a fixed module count.

\subsection{Additional Reference Coverage and Backend Configurations}
\label{app:comparison-coverage}

Of the 48 programs with 32 qubits, the fixed order and tree configurations of QuCLEAR complete 48 and 36, respectively, and direct canonical NOR decomposition completes 12. The six main and three additional references all complete on 12 programs. Of the 22 physical routing programs, the fixed order and tree configurations of QuCLEAR complete 17 and 16, respectively. The intersection of the six main references and these two additional references contains 48 program--seed pairs. CX counts for the additional references are compared only on the shared set of programs that complete under matching specifications.

The backend sensitivity check uses Qiskit 2.1.2 with \texttt{optimization\_level=1}, an all-to-all logical backend, and fixed random seed 20261101 to compile the source-level optimum, empty-cache, and full-cache candidates and the six fixed reference implementations under the same configuration. The comparison covers the deduplicated sets of 134 main evaluation programs and 72 independent confirmation programs.

\subsection{Per-Program Results for the Independent Confirmation Set}
\label{app:confirm-data}
Table~\ref{tab:instance-confirm} lists complete-circuit resources for each program in the independent confirmation set.

\begingroup
\scriptsize
\setlength{\tabcolsep}{4pt}
\renewcommand{\arraystretch}{1.05}
\begin{longtable}{rrllrrrr}
\caption{Per-program resources for 72 ordered programs on the 36 nonisomorphic graphs of the independent confirmation set}
\label{tab:instance-confirm}\\
\toprule $n$ & $d$ & Modules & ID/order & Ref. CX & Our CX & Our depth & Our 1Q\\\midrule\endfirsthead
\toprule $n$ & $d$ & Modules & ID/order & Ref. CX & Our CX & Our depth & Our 1Q\\\midrule\endhead
48 & 3 & Mixed & 0/A & 1,228 & 986 & 714 & 1,368\\
48 & 3 & Mixed & 0/S & 1,338 & 1,030 & 704 & 1,368\\
48 & 3 & Mixed & 1/A & 1,198 & 988 & 752 & 1,368\\
48 & 3 & Mixed & 1/S & 1,370 & 1,032 & 665 & 1,368\\
48 & 3 & Two-vertex & 0/A & 1,130 & 880 & 480 & 1,368\\
48 & 3 & Two-vertex & 0/S & 1,216 & 954 & 564 & 1,368\\
48 & 3 & Two-vertex & 1/A & 1,134 & 886 & 502 & 1,368\\
48 & 3 & Two-vertex & 1/S & 1,206 & 946 & 553 & 1,368\\
48 & 5 & Mixed & 0/A & 2,324 & 2,108 & 3,348 & 3,972\\
48 & 5 & Mixed & 0/S & 2,572 & 2,148 & 2,763 & 4,046\\
48 & 5 & Mixed & 1/A & 2,352 & 2,104 & 3,486 & 3,970\\
48 & 5 & Mixed & 1/S & 2,546 & 2,152 & 2,688 & 4,032\\
48 & 5 & Two-vertex & 0/A & 2,278 & 2,052 & 3,269 & 4,000\\
48 & 5 & Two-vertex & 0/S & 2,352 & 2,118 & 2,623 & 4,056\\
48 & 5 & Two-vertex & 1/A & 2,270 & 2,048 & 3,208 & 3,996\\
48 & 5 & Two-vertex & 1/S & 2,360 & 2,124 & 2,556 & 4,084\\
48 & 7 & Mixed & 0/A & 4,244 & 3,260 & 6,633 & 6,884\\
48 & 7 & Mixed & 0/S & 4,538 & 3,308 & 6,205 & 6,994\\
48 & 7 & Mixed & 1/A & 4,272 & 3,258 & 7,087 & 6,870\\
48 & 7 & Mixed & 1/S & 4,516 & 3,310 & 6,193 & 6,984\\
48 & 7 & Two-vertex & 0/A & 4,178 & 3,206 & 6,860 & 6,896\\
48 & 7 & Two-vertex & 0/S & 4,256 & 3,268 & 5,710 & 6,964\\
48 & 7 & Two-vertex & 1/A & 4,184 & 3,204 & 6,743 & 6,908\\
48 & 7 & Two-vertex & 1/S & 4,292 & 3,292 & 6,381 & 7,016\\
96 & 3 & Mixed & 0/A & 2,426 & 1,938 & 1,018 & 2,736\\
96 & 3 & Mixed & 0/S & 2,786 & 2,018 & 639 & 2,736\\
96 & 3 & Mixed & 1/A & 2,378 & 1,976 & 1,402 & 2,736\\
96 & 3 & Mixed & 1/S & 2,730 & 2,036 & 721 & 2,736\\
96 & 3 & Two-vertex & 0/A & 2,266 & 1,768 & 665 & 2,736\\
96 & 3 & Two-vertex & 0/S & 2,406 & 1,904 & 790 & 2,736\\
96 & 3 & Two-vertex & 1/A & 2,246 & 1,780 & 560 & 2,736\\
96 & 3 & Two-vertex & 1/S & 2,422 & 1,922 & 614 & 2,736\\
96 & 5 & Mixed & 0/A & 4,704 & 4,212 & 5,292 & 7,958\\
96 & 5 & Mixed & 0/S & 5,102 & 4,280 & 3,458 & 8,132\\
96 & 5 & Mixed & 1/A & 4,622 & 4,208 & 4,035 & 7,938\\
96 & 5 & Mixed & 1/S & 5,068 & 4,294 & 3,098 & 8,092\\
96 & 5 & Two-vertex & 0/A & 4,548 & 4,094 & 4,317 & 7,992\\
96 & 5 & Two-vertex & 0/S & 4,736 & 4,262 & 3,629 & 8,148\\
96 & 5 & Two-vertex & 1/A & 4,514 & 4,102 & 3,959 & 7,988\\
96 & 5 & Two-vertex & 1/S & 4,674 & 4,244 & 3,490 & 8,124\\
96 & 7 & Mixed & 0/A & 8,512 & 6,518 & 11,456 & 13,750\\
96 & 7 & Mixed & 0/S & 8,972 & 6,600 & 9,148 & 13,904\\
96 & 7 & Mixed & 1/A & 8,536 & 6,522 & 11,285 & 13,750\\
96 & 7 & Mixed & 1/S & 9,024 & 6,602 & 8,987 & 13,976\\
96 & 7 & Two-vertex & 0/A & 8,366 & 6,410 & 9,446 & 13,824\\
96 & 7 & Two-vertex & 0/S & 8,566 & 6,590 & 8,593 & 14,052\\
96 & 7 & Two-vertex & 1/A & 8,378 & 6,406 & 11,026 & 13,812\\
96 & 7 & Two-vertex & 1/S & 8,592 & 6,584 & 9,235 & 14,048\\
192 & 3 & Mixed & 0/A & 4,762 & 3,996 & 1,537 & 5,472\\
192 & 3 & Mixed & 0/S & 5,338 & 4,136 & 778 & 5,472\\
192 & 3 & Mixed & 1/A & 4,744 & 3,954 & 1,075 & 5,472\\
192 & 3 & Mixed & 1/S & 5,414 & 4,100 & 737 & 5,472\\
192 & 3 & Two-vertex & 0/A & 4,512 & 3,546 & 742 & 5,472\\
192 & 3 & Two-vertex & 0/S & 4,830 & 3,816 & 708 & 5,472\\
192 & 3 & Two-vertex & 1/A & 4,514 & 3,538 & 704 & 5,472\\
192 & 3 & Two-vertex & 1/S & 4,898 & 3,862 & 730 & 5,472\\
192 & 5 & Mixed & 0/A & 9,380 & 8,420 & 5,301 & 15,874\\
192 & 5 & Mixed & 0/S & 10,158 & 8,572 & 4,423 & 16,112\\
192 & 5 & Mixed & 1/A & 9,368 & 8,422 & 5,608 & 15,884\\
192 & 5 & Mixed & 1/S & 10,074 & 8,564 & 4,206 & 16,158\\
192 & 5 & Two-vertex & 0/A & 9,084 & 8,192 & 4,769 & 15,996\\
192 & 5 & Two-vertex & 0/S & 9,472 & 8,524 & 3,807 & 16,284\\
192 & 5 & Two-vertex & 1/A & 9,090 & 8,190 & 5,148 & 15,984\\
192 & 5 & Two-vertex & 1/S & 9,464 & 8,494 & 4,322 & 16,268\\
192 & 7 & Mixed & 0/A & 17,012 & 13,034 & 15,435 & 27,504\\
192 & 7 & Mixed & 0/S & 18,014 & 13,218 & 11,588 & 27,942\\
192 & 7 & Mixed & 1/A & 16,972 & 13,038 & 15,219 & 27,518\\
192 & 7 & Mixed & 1/S & 17,824 & 13,200 & 10,335 & 27,856\\
192 & 7 & Two-vertex & 0/A & 16,752 & 12,812 & 13,809 & 27,632\\
192 & 7 & Two-vertex & 0/S & 17,140 & 13,144 & 11,720 & 28,044\\
192 & 7 & Two-vertex & 1/A & 16,736 & 12,816 & 13,915 & 27,616\\
192 & 7 & Two-vertex & 1/S & 17,088 & 13,130 & 10,339 & 27,984\\
\bottomrule\end{longtable}
\endgroup
\LongTableNote{ID is the generation repetition index. A denotes ascending order and S denotes the fixed pseudorandom target order.}

\subsection{Paired Results for Two Target Orders on Public Graphs}
\label{app:public-data}
Table~\ref{tab:instance-public} summarizes the resources of the 24 ordered programs by graph. The values before and after each slash correspond to ascending order (A) and the fixed pseudorandom order (S), respectively. Each order is measured as a separate program.

\begingroup
\scriptsize
\setlength{\tabcolsep}{4pt}
\renewcommand{\arraystretch}{1.05}
\begin{longtable}{@{}l r r r r r@{}}
\caption{Resources for 24 programs on 12 public graphs under two target orders}
\label{tab:instance-public}\\
\toprule Graph & $n$ & \shortstack{Ref. CX\\A/S} & \shortstack{Our CX\\A/S} & \shortstack{Our depth\\A/S} & \shortstack{Param. rotations\\A/S}\\\midrule\endfirsthead
\toprule Graph & $n$ & \shortstack{Ref. CX\\A/S} & \shortstack{Our CX\\A/S} & \shortstack{Our depth\\A/S} & \shortstack{Param. rotations\\A/S}\\\midrule\endhead
david.col & 87 & 10,088/10,072 & 6,780/6,772 & 13,615/13,453 & 1,274/1,274\\
games120.col & 120 & 16,228/16,226 & 12,326/12,330 & 12,561/12,951 & 2,039/2,039\\
huck.col & 74 & 5,652/5,644 & 3,432/3,420 & 5,677/5,910 & 1,057/1,057\\
jean.col & 80 & 5,222/5,234 & 3,500/3,512 & 6,138/6,456 & 908/908\\
myciel3.col & 11 & 260/260 & 226/226 & 397/417 & 187/187\\
myciel4.col & 23 & 1,604/1,604 & 1,152/1,152 & 2,107/2,428 & 391/391\\
myciel5.col & 47 & 8,272/8,272 & 4,536/4,536 & 9,000/9,879 & 799/799\\
myciel6.col & 95 & 19,828/19,828 & 15,840/15,840 & 32,164/34,439 & 1,615/1,615\\
queen5\_5.col & 25 & 3,240/3,240 & 3,240/3,240 & 7,162/7,612 & 425/425\\
queen6\_6.col & 36 & 6,096/6,096 & 6,096/6,096 & 13,388/14,291 & 612/612\\
queen7\_7.col & 49 & 10,248/10,248 & 10,248/10,248 & 21,830/23,374 & 833/833\\
queen8\_8.col & 64 & 15,936/15,936 & 15,936/15,936 & 34,076/36,318 & 1,088/1,088\\
\bottomrule\end{longtable}
\endgroup
\LongTableNote{The reference is the per-program CX envelope of open CPG and the two matched exact kernels. A/S values are shown in the same order.}

\subsection{CX Overhead of Physical Routing and Terminal Wire-Order Restoration}
\label{app:routing-ledger}

Table~\ref{tab:routing-ledger} summarizes the routing input CX counts, final physical CX counts, and terminal wire-order restoration costs for the linear topology and two-dimensional grid. The additional restoration CX count measures the CX gates added by decomposing the SWAP gates needed to restore the original wire order, before the final local simplification pass. The net increase in physical CX is the difference between the final physical circuit and the routing input circuit, reflecting the combined effects of physical mapping, terminal wire-order restoration, and local simplification. The routing inputs have already undergone common native synthesis.

\begin{table}[H]
\centering
\caption{CX resources for physical routing and terminal wire-order restoration}
\label{tab:routing-ledger}

\begin{threeparttable}
\scriptsize
\setlength{\tabcolsep}{4pt}
\renewcommand{\arraystretch}{1.05}

\begin{tabular*}{\textwidth}{@{\extracolsep{\fill}}lrrrr@{}}
\toprule
Topology & Routing input CX & Final physical CX & Added restoration CX & Net physical CX increase\\
\midrule
Linear & 104,876 & 1,781,104 & 166,149 & 1,676,228\\
Two-dimensional grid & 104,738 & 598,736 & 49,896 & 493,998\\
\bottomrule
\end{tabular*}

\begin{tablenotes}[flushleft]
\scriptsize
\item Each topology aggregates 66 program--seed pairs from 22 programs with three fixed seeds. Output circuits are compared with routing inputs for the same program and seed.
\end{tablenotes}
\end{threeparttable}
\end{table}

\subsection{Single-Candidate Compilation Time and Memory Overhead at a Fixed Module Count}
\label{app:single-path}

Table~\ref{tab:single-path} compares the time and memory costs of the full-cache and source-level optimal cache paths with eight true-twin classes, compiling one candidate per path. To isolate the effects of module size and module count, the experiment fixes the quotient graph at $K_{2,6}$ and uses $n=16,32,64,128,256$ data qubits. Each condition is run in three independent single-threaded Windows processes. The time metric is the sum of the median times over the three runs for each of five separate stages (structure identification, source-level selection, circuit construction, native gate set translation, and native circuit post-processing). It differs from the wall-clock time of one complete compilation. The RSS metric is the median of the sampled process peaks across the three runs.

\begin{table}[H]
\centering
\caption{Time and memory overhead of two single-candidate paths with eight fixed true-twin classes}
\label{tab:single-path}

\begin{threeparttable}
\scriptsize
\setlength{\tabcolsep}{4pt}
\renewcommand{\arraystretch}{1.05}

\begin{tabular*}{\textwidth}{@{\extracolsep{\fill}}rrrrr@{}}
\toprule
$n$ &
\shortstack{Full cache\\time} &
\shortstack{Source selection\\time} &
\shortstack{Full cache\\RSS} &
\shortstack{Source selection\\RSS}\\
\midrule
16 & 0.26 & 0.28 & 105.9 & 106.6\\
32 & 0.35 & 0.39 & 109.1 & 109.0\\
64 & 0.55 & 0.73 & 115.3 & 115.4\\
128 & 1.25 & 1.80 & 128.1 & 129.7\\
256 & 4.82 & 6.99 & 162.9 & 164.7\\
\bottomrule
\end{tabular*}

\begin{tablenotes}[flushleft]
\scriptsize
\item Time is the sum of the median execution times of the five stages (s). RSS is the median of the sampled peaks from three independent processes (MiB).
\end{tablenotes}
\end{threeparttable}
\end{table}

\end{document}